\documentclass[sigconf]{acmart}

\AtBeginDocument{%
  \providecommand\BibTeX{{%
    \normalfont B\kern-0.5em{\scshape i\kern-0.25em b}\kern-0.8em\TeX}}}

\copyrightyear{2024}
\acmYear{2024}
\setcopyright{acmlicensed}\acmConference[KDD '24]{Proceedings of the 30th ACM SIGKDD Conference on Knowledge Discovery and Data Mining}{August 25--29, 2024}{Barcelona, Spain}
\acmBooktitle{Proceedings of the 30th ACM SIGKDD Conference on Knowledge Discovery and Data Mining (KDD '24), August 25--29, 2024, Barcelona, Spain}
\acmDOI{10.1145/3637528.3671837}
\acmISBN{979-8-4007-0490-1/24/08}

\usepackage{color}
\usepackage{hyperref}
\usepackage[linesnumbered,ruled,vlined]{algorithm2e}
\usepackage{balance}
\usepackage{bibentry}
\usepackage{amsfonts}
\usepackage{epsfig}
\usepackage{epstopdf}
\usepackage{rotating}
\usepackage{caption}
\usepackage{float}
\usepackage{multicol}
\usepackage{graphicx}
\usepackage{bbding}
\usepackage{microtype}
\usepackage{graphicx}
\usepackage{subfigure}
\usepackage{booktabs} 
\usepackage{multirow}
\usepackage{amsmath}
\usepackage{subcaption}

\usepackage{mathtools}
\usepackage{etoolbox}
\usepackage{cases}
\usepackage{enumitem}
\usepackage{xcolor}
\usepackage{newtxmath}
\usepackage{setspace}
\usepackage{lipsum}
\usepackage{tabularx}
\usepackage{array}

\hypersetup{
    colorlinks=true,
    linkcolor=black,
    filecolor=black,      
    urlcolor=black,
    citecolor=black,
}
\newcommand{\red}[1]{\textcolor{red}{#1}}
\definecolor{brown}{RGB}{139,64,0}

\begin{document}

\title{
CheatAgent: Attacking LLM-Empowered Recommender \\ Systems via LLM Agent}

\author{Liang-bo Ning}
\authornote{Both authors contributed equally to this research.}
\affiliation{%
  \institution{The Hong Kong Polytechnic University}
  \city{Hong Kong}
  \country{China}
}
\email{BigLemon1123@gmail.com}

\author{Shijie Wang}
\authornotemark[1]
\affiliation{%
  \institution{The Hong Kong Polytechnic University}
  \city{Hong Kong}
  \country{China}
}
\email{shijie.wang@connect.polyu.hk}

\author{Wenqi Fan}
\authornote{Corresponding author: Wenqi Fan, Department of Computing, and 
 Department of Management and Marketing, The Hong Kong Polytechnic University.}
\affiliation{%
  \institution{The Hong Kong Polytechnic University}
  \city{Hong Kong}
  \country{China}
}
\email{wenqifan03@gmail.com}

\author{Qing Li}
\affiliation{%
  \institution{The Hong Kong Polytechnic University}
  \city{Hong Kong}
  \country{China}
}
\email{qing-prof.li@polyu.edu.hk}

\author{Xin Xu}
\affiliation{%
  \institution{The Hong Kong Polytechnic University}
  \city{Hong Kong}
  \country{China}
}
\email{xin.xu@polyu.edu.hk}

\author{Hao Chen}
\affiliation{%
  \institution{The Hong Kong Polytechnic University}
  \city{Hong Kong}
  \country{China}
}
\email{sundaychenhao@gmail.com}

\author{Feiran Huang}
\affiliation{%
  \institution{Jinan University}
  \city{Guangzhou}
  \country{China}
}
\email{huangfr@jnu.edu.cn}

\renewcommand{\shortauthors}{Liang-bo Ning et al.}


\begin{abstract}

Recently, Large Language Model (LLM)-empowered recommender systems (RecSys) have brought significant advances in personalized user experience and have attracted considerable attention.
Despite the impressive progress, the research question regarding the safety vulnerability of LLM-empowered RecSys still remains largely under-investigated. 
Given the security and privacy concerns, it is more practical to focus on attacking the black-box RecSys, where attackers can only observe the system’s inputs and outputs.
However,  traditional attack approaches employing reinforcement learning (RL) agents are not effective for attacking LLM-empowered RecSys due to the limited capabilities in processing complex textual inputs, planning, and reasoning. 
On the other hand, LLMs provide unprecedented opportunities to serve as attack agents to attack RecSys because of their impressive capability in simulating human-like decision-making processes. 
Therefore, in this paper, we propose a novel attack framework called \emph{CheatAgent} by harnessing the human-like capabilities of LLMs, where an LLM-based agent is developed to attack LLM-Empowered RecSys.
Specifically, our method first identifies the insertion position for maximum impact with minimal input modification.
After that, the LLM agent is designed to generate adversarial perturbations to insert at target positions.
To further improve the quality of generated perturbations, we utilize the prompt tuning technique to improve attacking strategies via feedback from the victim RecSys iteratively. 
Extensive experiments across three real-world datasets demonstrate the effectiveness of our proposed attacking method.

\end{abstract}


\maketitle

\section{Introduction}\label{sec:introduction}
Recommender Systems (\textbf{RecSys}) play a vital role in capturing users' interests and preferences across various fields~\cite{fan2019graph}, such as e-commerce (e.g., Amazon, Taobao), social media (e.g., Twitter, Facebook), etc. Traditional RecSys typically rely on users' historical interactions to analyze user behaviors and item characteristics~\cite{he2017neural}. 
Recent developments in deep learning (DL) have introduced neural networks like Graph Neural Networks (GNNs) and Recurrent Neural Networks (RNNs) in RecSys to further improve recommendation performance~\cite{he2020lightgcn,fan2019deep_dscf}.
Although DL-based methods effectively model the representations of users and items, they struggle with encoding textual information (e.g., item titles, user reviews) for reasoning on user's prediction~\cite{fan2023recommender,lin2023can}.
Recently, due to the powerful language understanding and in-context learning capabilities, Large Language Models (LLMs) have provided great potential to revolutionize RecSys~\cite{geng2022recommendation,sun2019bert4rec,bao2023tallrec}. 
For instance, P5~\cite{geng2022recommendation} leverages LLM's (i.e. T5~\cite{raffel2020exploring}) capabilities to significantly enhance recommendation performance by understanding nuanced user preferences and item descriptions. 
Despite the aforementioned success, there is a critical issue that remains largely unexplored: \textbf{the safety vulnerability of LLM-empowered recommender systems under adversarial attacks}, which hinders their adoption in various real-world applications, especially those high-stake environments like finance and healthcare.

\begin{figure}[t]
\setlength{\abovecaptionskip}{3mm}
\centering
\includegraphics[width=0.9\linewidth]{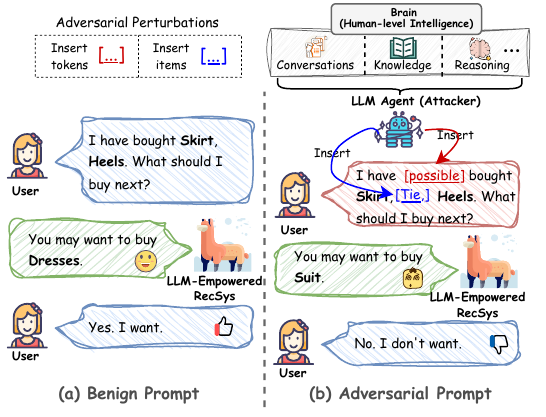}
\caption{The illustration of the adversarial attack for recommender systems in the era of LLMs. 
Attackers leverage the LLM agent to insert some tokens (e.g., words) or items in the user's prompt to manipulate the LLM-empowered recommender system to make incorrect decisions. 
}
\label{fig:Illustration}
\end{figure}

Given the need for security and privacy, a practical attacking strategy in black-box recommender systems involves utilizing reinforcement learning (RL) agents to conduct poisoning attacks~\cite{fan2021attacking,fan2023adversarial}.  
To be specific, under the black-box setting, attackers have no access to the models or parameters of the victim RecSys. Instead, they are limited to observing the system's inputs and outputs only.
For example, most existing solutions, such as KGAttack~\cite{chen2022knowledge}, PoisonRec~\cite{song2020poisonrec}, and CopyAttack~\cite{fan2021attacking},  develop RL-based agents to obtain malicious user profiles (i.e., a series of items) and inject them into the victim RecSys for manipulating system's decision.
Despite the impressive progress in attacking recommender systems under the black-box setting, most existing attack approaches still suffer from several limitations.  
\emph{First}, vanilla RL-based agents struggle with processing the textual input (e.g., item's title and descriptions) and context awareness, resulting in difficulty in attacking LLM-empowered RecSys which mainly takes text as input and generates relevant responses in natural language. 
\emph{Second}, due to the lack of a vast amount of open-world knowledge, most existing methods optimize the RL-based agent attackers from scratch without human-level intelligence, which subsequently leads to poor capability in planning and reasoning the attacking strategies under the black-box setting.   
Hence, it is desirable to design a novel paradigm for attacking black-box recommender systems in the era of LLMs.

More recently, Large Language Models (LLMs) have achieved great success in various fields, such as psychology~\cite{aher2023using}, drug discovery~\cite{li2023empowering}, and health~\cite{zhang2023huatuogpt}, demonstrating their remarkable potential in approximating human-level intelligence. 
This impressive capability is attributed to the training on vast textual corpora (i.e., open-world knowledge) with a huge amount of model parameters~\cite{fan2023recommender,zhao2023survey}. 
As such, LLMs can well comprehend human common sense in natural language and perform complex reasoning, so as to simulate human-like decision-making processes~\cite{wang2023survey}.
Given their advantages, LLMs provide unprecedented opportunities to overcome the limitations faced by current RL-based attack methods and serve as attack agents to attack RecSys. 
Therefore, in this work, we propose a novel attacking strategy to attack the LLM-empowered recommender systems by taking advantage of LLM as the autonomous agent for making human-like decisions.
As shown in Figure~\ref{fig:Illustration}, an LLM-based agent with human-like intelligence is introduced to generate an adversarial prompt by adding slight perturbations (e.g., words and items) on the original prompt,  so as to mislead LLM-empowered RecSys to make unsatisfactory recommendations.

In this paper, we propose a novel attack framework (\emph{CheatAgent}) to investigate the safety vulnerability of LLM-empowered RecSys under the black-box setting.
Specifically, an LLM is introduced as an intelligence agent to generate adversarial perturbations in users' prompts for attacking the LLM-based system.
To address the vast search space on insertion position and perturbation selection for the LLM agent, we first propose insertion positioning to identify the input position for maximum impact with minimal input modification. 
After that, LLM agent-empowered perturbation generation is proposed to generate adversarial perturbations to insert at target positions.
Due to the domain-specific knowledge gap between the attack agent and LLM-empowered RecSys, we further develop a self-reflection policy optimization to enhance the effectiveness of the attacks.
Our major contributions of this paper are as follows: 
\begin{itemize}
    \item We study a novel problem of whether the existing LLM-empowered recommender systems are robust to slight adversarial perturbations. To the best of our knowledge, this is the first work to investigate the safety vulnerability of the LLM-empowered recommender systems. 
    
    \item We introduce a novel strategy to attack black-box recommender systems in the era of LLMs, where an LLM-based agent is developed to generate adversarial perturbations on input prompts, so as to mislead LLM-empowered recommender systems for making incorrect decisions.

    \item We propose a novel framework \textbf{CheatAgent} to attack LLM-empowered recommender systems under the black-box setting via the LLM-based attack agent, which efficiently crafts imperceptible perturbations in users' prompt to perform effective attacks.

    \item We conduct extensive experiments on three real-world datasets to demonstrate the safety vulnerability of the LLM-empowered recommender systems against adversarial attacks and the attacking effectiveness of our proposed attack method.

\end{itemize}

\section{Problem Statement}
\subsection{Notation and Definitations}~\label{sec:preliminary}
The objective of RecSys is to understand users' preferences by modeling the interactions (\textit{e.g.}, clicks, purchases, etc.) between users $U = \{u_1, u_2, \cdots, u_{|U|}\}$ and items $V = \{v_1, v_2, \cdots, v_{|V|}\}$. 
Within the framework of a general LLM-empowered RecSys $Rec_{\Theta}$ with parameters $\Theta$, we denote an input-output sequence pair as $(X, Y)$, consisting of a recommendation prompt template $P=[x_1, x_2, \cdots, x_{|P|}]$, user  $u_i$, and the user's historical interactions towards items $V^{u_i}=[v_1, v_2, \cdots,v_{|V^{u_i}|} ]$ (\textit{i.e.}, user's profile). 
Based on the above definition, a typical input can be denoted as:

\centerline{$X = [P;u_i;V^{u_i}] =[x_1, \cdots,  \text{user}\_u_i, \cdots, \text{items}\_V^{u_i}, \cdots, x_{|P|} ]$. }

\noindent For instance, as shown in Figure~\ref{fig:framework}, a specific input-output pair with user-item interaction in the language model for recommendation can be represented as:

\centerline{$\begin{aligned} X = [ & \text{What, is, the, top, recommended, item, for, }  \emph{User\_637}, \text{ who,} \\ 
    & \text{has, interacted, with, }  \emph{item\_1009}, ..., \emph{item\_4045}, ? ], \end{aligned}$ }

\centerline{ $\begin{aligned} Y=&[\emph{item\_1072}], \end{aligned}$ }

\noindent where $u_i = [User\_637]$ and $V^{u_i}= [\emph{item\_1009}, ..., \emph{item\_4045}]$. The other tokens belong to the prompt template $P$. 

After that, LLM-empowered RecSys will generate recommendations based on the textual input. 
The auto-regressive language generation loss (\textit{i.e.}, Negative Log-Likelihood) is employed to 
evaluate the discrepancy between the predictions and the target output, 
defined as follows:

\centerline{$\mathcal{L}_{Rec} (X, Y) =   \frac{1}{|Y|} { \sum_{t=1}^{|Y|}} -\log p(Y_t|{X}, Y_{<t}),$}

\noindent 
where $p(Y_t|{X}, Y_{<t})$ represents the probability assigned to the item that users are interested in. 
Small $\mathcal{L}_{Rec} (X, Y)$ indicates that RecSys can accurately predict the target label $Y$ and vice versa.

\subsection{Attacker's Capabilities} 
In this work, we will focus on attacking black-box LLM-empowered recommender systems, where inherent details of the victim LLM-empowered recommender system, including architectures, gradients, parameters, etc., are restricted from access. 
In other words, the attackers can devise adversarial perturbations by solely querying the target system and observing the resulting output probabilities, similar to the soft-label black-box setting in~\cite{liu2023hqa,jin2020bert}.

\subsection{Attacker's Objective} \label{sec:attack_objective}
The overall objective of attackers is to conduct \emph{untargeted attacks} by undermining the overall performance of the victim LLM-empowered RecSys, specifically by causing the target RecSys to prioritize \emph{irrelevant items} that are of no interest to users. 
Note that these malicious manipulations can undermine the overall user experience and compromise the trustworthiness of RecSys. 
More specifically, to generate incorrect recommendations for user $u_i$, attackers aim to carefully craft adversarial perturbations and insert them into the input $X=[P;u_i; V^{u_i}]$ as $\hat{X}= \mathbb{I}(X, \delta | s)$ to deceive the victim RecSys to learn the users' preference, where $\mathbb{I}(X, \delta | s)$ represent to insert perturbation $\delta$ at the position $s$ of the input $X$. 
In the context of LLM-based recommender systems, two operations can be designed for attackers to generate adversarial perturbations on input: 1) insert the tailored perturbations into the prompt template (i.e., $\hat X = [\hat P;u_i; V^{u_i}] = [\mathbb{I}(P, \delta | s);u_i; V^{u_i}]$), and 2) perturb the users' profiles to distort their original preference (i.e., $\hat X = [P;u_i; \hat{V}^{u_i}] = [P;u_i; \mathbb{I}(V^{u_i}, \delta | s)]$).

Given these two different attacking operations, adversarial perturbations applied to the recommendation prompt $P$ and users' profiles $V^{u_i}$ differ in nature. 
Specifically, words or characters can be used as perturbations inserted into the recommendation prompt  $P$, while items serve as perturbations inserted into user profiles $V^{u_i}$.
For the simplicity of notation, $\delta$ is employed to uniformly represent these two forms of perturbations.  
Mathematically, adversarial perturbations $\delta$ can be generated by decreasing the recommendation performance, and the overall objective is formulated as follows: 

\centerline{$\delta = \underset{\delta: \| \hat X - X \|_0 \le  {\triangle}}{\arg \max} ~\mathcal{L}_{Rec}(\hat X, Y),$}

\noindent
where $\| \hat X - X \|_0$ is the Hamming distance between the benign input and adversarial input~\cite{zhang2023certified} and the $\triangle$ is the predefined upper bound to constrain the magnitude of perturbations.

\begin{figure*}[t]
\setlength{\abovecaptionskip}{3mm}
\centering
\includegraphics[width=1\linewidth]{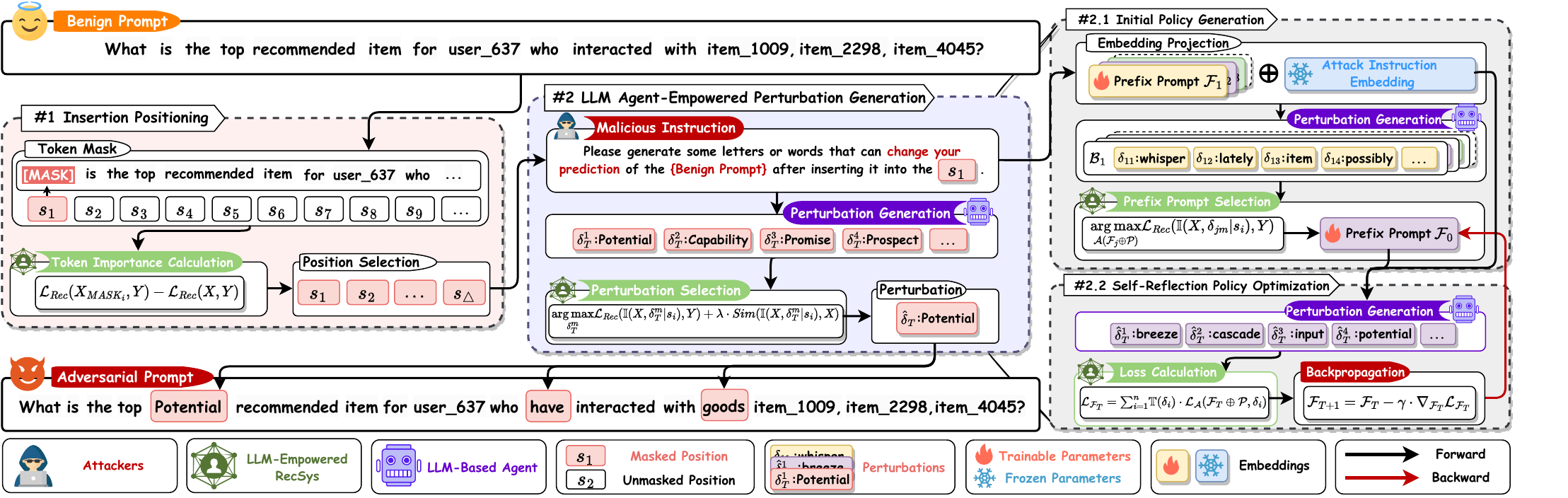}
\caption{The overall framework of the proposed CheatAgent. 
Insertion positioning first locates the token with the maximum impact. Then, LLM agent-empowered perturbation generation is proposed to leverage the LLM as the attacker agent to generate adversarial perturbations. It contains two processes: 1) Initial Policy Generation searches for a great attack policy initialization, and 2) Self-Reflection Policy Optimization fine-tunes the prefix prompt to update the attack policy of the LLM-based agent. 
}
\label{fig:framework}
\end{figure*}

\section{Methodology}\label{sec:method}

\subsection{An Overview of the Proposed CheatAgent}
In order to conduct black-box attacks on target LLM-empowered RecSys, adversarial perturbations are generated to modify the input prompts to mislead the generation of LLM-empowered systems. 
To achieve this goal, we propose a novel attacking strategy, in which an LLM-based agent (attacker) is developed to effectively craft input prompts, due to the powerful language comprehension, reasoning abilities, and rich open-world knowledge of LLMs.
However, developing malicious LLM-based agents to perform attacks under the black-box setting faces challenges due to numerous options for both insertion positions and perturbation selection.

To address these challenges, we propose a novel framework (\emph{CheatAgent}), which utilizes the prompt tuning techniques to learn attacking strategies and generate high-quality adversarial perturbations via interactions with the victim RecSys iteratively.
As illustrated in Figure~\ref{fig:framework}, the overall framework of our proposed method consists of two main components: Insertion Positioning and LLM Agent-Empowered Perturbation Generation.
First, we aim to position the inserting tokens to achieve maximum impact with minimal input modification.
Specifically, we identify the tokens within the prompt that possess the substantial impact to deceive the victim model by employing minimal perturbations.
Second, LLM agent-empowered perturbation generation is proposed to fully leverage the powerful capabilities of LLMs in comprehending and generating natural language, as well as reasoning with open-world knowledge to generate adversarial perturbations to deceive the target system. 
The proposed approach contains two processes: initial policy generation and self-reflection policy optimization. 
These two processes initialize and fine-tune the attack policy based on the feedback from the target system by utilizing prompt tuning techniques to perform effective attacks.

\subsection{Insertion Positioning}
As the impact of each token within the prompt can vary significantly, positioning the insertion tokens is crucial for conducting effective attacks~\cite{gao2018black,garg2020bae}. 
Consequently, we propose to insert new tokens adjacent to the tokens that contribute more towards the final prediction and can achieve maximum impact with minimal input modification. 
Therefore, we first evaluate the importance of each word/item within the input $X$ and locate the token with the maximum impact. 
As shown in the first component of Figure~\ref{fig:framework}, we propose to mask the token from the input sentence and then examine the change it brings to the final predictions, so as to evaluate the token importance of the input prompt. 
Given an input $X$ containing $|X|$ tokens, masking a specific $i$-th token from the input $X$ can be defined as follows: ${X}_{MASK_i} = \mathbb{I}(X, [MASK]|i)$. 
The significance of the $i$-th token is determined by measuring the variation in prediction performance between the original input and the input with the token masked, as follows:

\centerline{$I_i = \mathcal{L}_{Rec}{({X}_{MASK_i}, Y)} - \mathcal{L}_{Rec}{({X}, Y)}.$}

After calculating the importance for $|X|$ tokens respectively, we can obtain the importance list $[I_1, I_2, ..., I_{|X|}]$. 
Then, a position list is generated by selecting the tokens with top-$\triangle$ importance scores, defined by: $\mathcal{S} = [s_1, s_2, \cdots, s_\triangle].$

\subsection{LLM Agent-Empowered Perturbation Generation}\label{sec:promt_tuning}
Once the tokens with the highest impact have been identified, the next crucial step is to determine the perturbations to be inserted. 
Due to the superiority of the LLM-based agent in comprehending natural language and its abundant knowledge derived from abundant training data, we propose an LLM-based agent paradigm to attack LLM-empowered RecSys, where an auxiliary large language model is designed as the attack agent to generate high-quality perturbations for the specific positions. 
However, manipulating the target RecSys needs to select the most effective token as an adversarial perturbation from a vast collection of options, which is a highly complex and challenging task. 
Direct utilization of adversarial perturbations generated by the LLM-based agent based on the initial attack policy often fails to achieve the desired attack performance due to the lack of domain-specific knowledge. 
Moreover, due to the extensive number of internal parameters in the LLM, it is impractical and inefficient to fine-tune the entire LLM agent by interacting with the target  RecSys.

To address these challenges, as shown in Figure~\ref{fig:framework}, we propose a prompt tuning-based attack policy optimization strategy, 
in which a trainable prefix prompt $\mathcal{F}$ is designed to integrate into the attackers' instruction  $\mathcal{P}$ in the embedding space. 
Meanwhile, we only fine-tune the prefix prompt $\mathcal{F}$ by interacting with the target RecSys to optimize the attack policy of the LLM-based agent.
Given that the task performance of large language models is significantly influenced by the quality of the input prompts~\cite{zhang2022automatic}, freezing the parameters of the LLM-based agent results in the attack policy being highly dependent on the input instruction provided by attackers.
Therefore, the LLM-based agent can adjust the attack policy by fine-tuning the task-specific instruction given by attackers, thereby effectively reducing the computational burden and time consumption of retraining the entire LLM.

The proposed method in this component is comprised of two main steps: 1) Initial Policy Generation, and 2) Self-Reflection Policy Optimization. 
To be specific, Initial Policy Generation aims to search for an appropriate prefix prompt to initialize a benchmark attack policy to minimize subsequent iterations for policy tuning.
Then, given the initialized prefix prompt, we propose a self-reflection policy optimization strategy to fine-tune the prefix prompt and update the attack policy of the LLM-based agent by utilizing the feedback from the victim RecSys.

\subsubsection{\textbf{Initial Policy Generation}}
Before updating the attack policy by fine-tuning the trainable prefix prompt, the agent must generate an initial policy to start optimization.
Poor initialization can lead the agent to get stuck in local optimal when learning the attack policy~\cite{daniely2016toward}, bringing difficulties in effectively attacking the target system.  
Therefore, to enhance the attack performance of the generated perturbations and decrease the number of subsequent policy tuning iterations, we propose to search for an appropriate prefix prompt to initialize the attack policy in the LLM-based attacker agent. 
To achieve this goal, we randomly initialize multiple prefix prompts and combine them with the attack's instructions respectively to generate multiple adversarial perturbations. 
Each perturbation is evaluated for its attack performance, and the prefix prompt that can generate the perturbation with the greatest impact in misleading the target RecSys is deemed the optimal initialization.

We use $\mathcal{P} \in \{\mathcal{P}_P, \mathcal{P}_{V^{u_i}}\}$ to represent the attacker's instructions, which is exploited to guide the LLM-based agent to generate perturbations. 
As we mentioned in Section~\ref{sec:attack_objective}, $\delta$ has two forms of adversarial perturbations in attacking LLM-empowered RecSys, so distinct instructions $\mathcal{P}_P$ and $\mathcal{P}_{V^{u_i}}$ are employed to generate perturbations that are inserted to the prompt $P$ and users' profiles $V^{u_i}$ (more details about the instructions given by attackers are shown in Table~\ref{tab:generation_prompt} of \textbf{Appendix}~\ref{appendix:implementation}). 
Technically, we first initialize $k$ prefix prompts $[\mathcal{F}_1, ..., \mathcal{F}_k]$, each prefix is combined with the attacker's instruction $\mathcal{P}$ in the embedding space and fed into the LLM-based agent $\mathcal{A}$ to generate $n$ perturbation candidates, defined by:
\begin{align}
\begin{split}
    \mathcal{B}_j = \mathcal{A}(\mathcal{F}_j \oplus \mathcal{P}),
\end{split}
\label{eq:initialization_candidate_generation}
\end{align}
where $\oplus$ is the combination operator and $\mathcal{B}_j=[\delta_{j1}, \delta_{j2}, ..., \delta_{jn}], j\in \{1,k\}$ is the perturbation candidates generated by the LLM-based agent $\mathcal{A}$ based on the combined prompt $\mathcal{F}_j \oplus \mathcal{P}$. 
After that, each perturbation candidate of $\mathcal{B}_j$ is iteratively inserted into the prompt $X$ at the position $s_i$. 
The perturbation that maximally undermines the prediction performance of the victim system is selected from all candidates, and the prefix used to generate this perturbation is considered as the initial prefix $\mathcal{F}_0$, defined by: 
\begin{align}
    \begin{split}
        \mathcal{F}_0 &= \underset{\mathcal{A}(\mathcal{F}_j \oplus \mathcal{P})}{\arg \max} \mathcal{L}_{Rec}(\mathbb{I}(X, \delta_{jm}|s_i), Y), j \in \{1,k\}, m \in \{1,n\}. \\
    \end{split}
\label{eq:selection}
\end{align}

Here we use $\mathcal{L}_{Rec}^{max} = {\max} \mathcal{L}_{Rec}(\mathbb{I}(X, \delta_{jm}|s_i),Y)$ to denote the maximum loss after inserting all candidates at position $s_i$ respectively, where $j \in \{1,k\}$ and $m \in \{1,n\}$. 

\subsubsection{\textbf{Self-Reflection Policy Optimization}}
Due to the domain-specific knowledge gap between the attack agent and the LLM-empowered RecSys that may be fine-tuned on the recommendation data, the initial attack policy based on the given prefix prompt can be sub-optimal.
To further optimize the attack policy and enhance the attack performance,
it is necessary to fine-tune the initialized prefix prompt $\mathcal{F}_0$ in LLM-based agent via the feedback (i.e., output) from the victim system under the black-box setting.
Specifically, we propose a black-box self-reflection prompt tuning strategy, which aims to determine the optimization direction according to the feedback produced by the target RecSys. 
First, the perturbations $\mathcal{B}_0=[\delta_1,..., \delta_n]$ generated by $\mathcal{A}(\mathcal{F}_0 \oplus \mathcal{P})$ are divided positive and negative categories.
Subsequently, we optimize the attack policy in a direction that enables the LLM-based agent to generate a higher number of positive perturbations, while minimizing the production of negative perturbations it generates. 
As the overall objective is to maximize $\mathcal{L}_{Rec}(\hat X, Y)$, by evaluating the effect of the perturbation on attack loss, we can classify perturbations into positive and negative, defined by: $\mathbb{T}(\delta_i)$, where $\mathbb{T}$ is an indicator function: 
\begin{align}
\label{eq:selector}
    \begin{split}
        \mathbb{T}(\delta_i) = 
        \begin{cases}
          1, & \text{ if } \mathcal{L}_{Rec}(\mathbb{I}(X, \delta_{j}|s_i),Y) \ge \mathcal{L}_{Rec}^{max},\\
          -1, & \text{ if } \mathcal{L}_{Rec}(\mathbb{I}(X, \delta_{j}|s_i),Y) < \mathcal{L}_{Rec}^{max},
        \end{cases}
    \end{split}
\end{align}
where $\mathbb{T}(\delta_i) = 1$ means $\delta_i$ can further enhance the attack performance, and it is considered as the positive perturbation. 
If $\delta_i$ is a negative perturbation, we compute the gradient of $\delta_i$ with respect to $\mathcal{F}_0$ and update $\mathcal{F}_0$ in the direction of gradient ascent. This ensures that $\mathcal{F}_0\oplus\mathcal{P}$ minimally guides the LLM to generate negative perturbations.
Based on the above definition, we can formulate the optimization problem as follows:
\vskip -0.2in
\begin{align}
\label{eq:prefix}
    \begin{split}
        \mathcal{L}_{\mathcal{F}_0} &= \sum_{i=1}^{n} \mathbb{T}(\delta_i) \cdot \mathcal{L}_{\mathcal{A}} (\mathcal{F}_0 \oplus \mathcal{P}, \delta_i) \\
        &= \sum_{i=1}^{n_+} \mathcal{L}_{\mathcal{A}} (\mathcal{F}_0 \oplus \mathcal{P}, \delta_i^+) - \sum_{j=1}^{n_{-}} \mathcal{L}_{\mathcal{A}} (\mathcal{F}_0 \oplus \mathcal{P}, \delta_j^-), 
    \end{split}
\end{align}
\vskip -0.1in
\noindent where $\mathcal{L}_{\mathcal{A}} (\mathcal{F}_0 \oplus \mathcal{P}, \delta_i) =   \frac{1}{|\delta_i|} {\sum_{t=1}^{|\delta_i|}} -\log p(\delta_i^t|{\mathcal{F}_0 \oplus \mathcal{P}}, \delta_i^{<t})$ is the negative log-likelihood loss. $n_+$ and $n_-$ are the number of positive perturbations $\delta_i^+$ and negative perturbations $\delta_j^-$, respectively. 
Minimizing Eq~\eqref{eq:prefix} promotes the LLM-based agent $\mathcal{A}$ to update its attack policy to generate more positive perturbations with a significant impact on the manipulation of target system's predictions. 
The optimization process is defined by: $\mathcal{F}_{T} =  \mathcal{F}_{T-1} - \gamma \cdot \nabla_{\mathcal{F}_{T-1}} \mathcal{L}_{\mathcal{F}_{T-1}},$
where $\gamma = 0.1$ is the learning rate and $T \in \{1,5\}$ is the number of policy optimization iterations. 

\subsubsection{\textbf{Final Perturbation Selection}}
Through backpropagation, we can obtain an optimized prefix prompt $\mathcal{F}_T$ that equips the LLM-based agent $\mathcal{A}$ with the powerful attack policy to generate high-quality perturbations $\mathcal{B}_T=[\delta_T^1, ..., \delta_T^n]$. 
Finally, the perturbation ${\hat \delta}_T$, which can not only induce the largest decrease in the performance of the target RecSys but also preserve high semantic similarity, is considered the optimal solution and inserted into the input prompt $X$. The optimal perturbation selection process is defined by:
\begin{align}
\label{eq:final_perturbation}
        {\hat \delta}_T =  \underset{\delta_T^m}{\arg \max} \mathcal{L}_{Rec}(\mathbb{I}(X, \delta_T^m|s_i), Y) 
         + \lambda \cdot Sim(\mathbb{I}(X, \delta_T^m|s_i), X), 
\end{align}
where $Sim(\mathbb{I}(X, \delta_T^m|s_i), X)$ is the cosine similarity between the perturbed prompt $\mathbb{I}(X, \delta_T^m|s_i)$ and the benign prompt $X$, and $\lambda=0.01$ is the hyper-parameter to balance the impact of these two aspects. 
The semantic similarity is computed by introducing an additional embedding model \emph{bge-large-en}~\cite{xiao2023c}. 
The whole process of the proposed CheatAgent is shown in \textbf{Algorithm}~\ref{al:CheatRec} (\textbf{Appendix}~\ref{appendix:pseudo}).

\section{Experiments}\label{sec:exp}
In this section, comprehensive experiments are conducted to demonstrate the effectiveness of the proposed method. 
Due to the space limitation, some details of the experiments and discussions are shown in \textbf{Appendix}~\ref{appendix:experimental_details} and \textbf{Appendix}~\ref{appendix:discussion}. 

\subsection{Experimental Details}

\subsubsection{\textbf{Datasets}. }
All experiments are conducted on three commonly-used datasets in RecSys: Movielens-1M (\textbf{ML1M})~\cite{movielens}, \textbf{Taobao}~\cite{taobao1}, and \textbf{LastFM}~\cite{xu2023openp5} datasets. 
The \textbf{ML1M} dataset provides movie ratings and user information, the \textbf{Taobao} dataset contains e-commerce transaction data, and the \textbf{LastFM} dataset offers user listening histories and music information.
The details of these datasets are summarised in \textbf{Appendix}~\ref{appendix:datasets}.

\subsubsection{\textbf{Victim LLM-based Recommender Systems}. }
\textbf{P5}~\cite{geng2022recommendation} and \textbf{TALLRec}~\cite{bao2023tallrec} are exploited to investigate the safety vulnerability of LLM-empowered recommender systems:
\begin{itemize}
    \item \textbf{P5} first converts all data, including user-item interactions, user descriptions, etc., to natural language sequences. It proposes several item indexing strategies, introduces the whole-word embedding to represent items, and fine-tunes the T5~\cite{raffel2020exploring} to improve the recommendation performance. 
    \item \textbf{TALLRec} transfers the recommendation problem to a binary textual classification problem. It fine-tunes the LLaMA~\cite{touvron2023llama} on the recommendation task and utilizes the user’s interaction history to forecast their interest in a forthcoming item by integrating item titles into a pre-defined prompt. 
\end{itemize}

\subsubsection{\textbf{Baselines}. } 
Multiple baselines are employed to investigate the vulnerability of the LLM-empowered RecSys, shown as follows: 
\begin{itemize}
\item \textbf{MD} manually designs an adversarial prompt with the opposite semantic meaning to the original prompt $X$ by inserting "not".
The used prompt is shown in \textbf{Appendix}~\ref{appendix:implementation} Table~\ref{tab:md}.
\item \textbf{RL}~\cite{fan2023untargeted} uses the Proximal Policy Optimization (PPO)~\cite{schulman2017proximal} to train the attack policy to generate adversarial perturbations. 
\item \textbf{GA}~\citep{lapid2023open} employs the genetic algorithm to find the adversarial perturbation and insert them to the end of the benign input.
\item \textbf{BAE}~\cite{garg2020bae} masks the crucial words within the input prompt and exploits the language model, i.e., BERT ~\citep{kenton2019bert}, to predict the contextually appropriate perturbations. 
\item \textbf{LLMBA}~\cite{xu2023llm} directly utilizes large language models to generate adversarial perturbations and insert them to the end of the benign input.
The prompts used for perturbation generation are shown in Table~\ref{tab:generation_prompt} of \textbf{Appendix}~\ref{appendix:implementation}.
\item \textbf{RP} selects items randomly from the item set and inserts them at a random position in users' profiles. 
\item \textbf{RT} selects words randomly from the vocabulary and inserts them at a random position in the benign prompt.
\item \textbf{RPGP} selects tokens randomly and inserts them at the position specified by the proposed method. 
\item \textbf{C-w/o PT} directly uses prompts to guide the LLM-based agent to generate perturbations without policy tuning. 
\item \textbf{CheatAgent} uses prompt-tuning to guide the LLM-based agent to produce high-quality perturbations.
\end{itemize}

\subsubsection{\textbf{Implementation}. }~\label{sec:implementation}
The proposed methods and all baselines are implemented by Pytorch. 
All victim models (\textbf{P5} and \textbf{TALLRec}) are implemented according to their official codes. 
For \textbf{P5} model, we use two different item indexing methods (i.e., random indexing and sequential indexing) to demonstrate the robustness of the generated adversarial perturbations. 
For \textbf{TALLRec} model, since it needs ratings to divide the user-interested items and user-hated items, we fine-tune the LLaMA model on a textual dataset reconstructed by \textbf{ML1M} dataset and test its vulnerability on this dataset. 

We initialize the population with a quantity of 50 and iterate for 10 epochs to obtain the final perturbation for \textbf{GA}. 
Bert~\cite{kenton2019bert} is used to generate 50 candidates, and \textbf{BAE} selects the perturbation that is most effective in undermining the recommendation performance. 
As for the proposed CheatAgent, we use distinct prompts $\mathcal{P} \in \{\mathcal{P}_P, \mathcal{P}_{V^{u_i}}\}$ to generate candidates as mentioned in Section~\ref{sec:attack_objective}. 
The prompts used for perturbation generation are shown in Table~\ref{tab:generation_prompt} of \textbf{Appendix}~\ref{appendix:implementation}. 
For \textbf{P5}, we set $k=10$ and $n=10$ as defaults, and for \textbf{TALLRec}, we set $k=6$ and $n=12$. 
\textbf{T5}~\cite{raffel2020exploring} is employed as the LLM-based agent $\mathcal{A}$. 
$\triangle$ is set to 3 for all methods, which means we can only insert three perturbed words/items into the input prompt $X$. 
Besides, during experiments, for the item within the user's profile $V_{u_i}$,  we observe that masking a pair of items and inserting perturbations to the middle of the maximum-impact items can achieve better attack performance. 
We argue that this may be due to the significant impact of the order of item interactions on user preferences. More experiments and discussion about this phenomenon are shown in Table~\ref{tab:insertion_position} of \textbf{Appendix}~\ref{appendix:experiments}.

\subsubsection{\textbf{Evaluation Metrics}. } 
For \textbf{P5} model, we consider two metrics, formulated as $\text{ASR-H}@r = 1 - \widehat{\text{H}@r}/{\text{H}@r}$ and $\text{ASR-N}@r = 1 - \widehat{\text{N}@r}/{\text{N}@r}$.
$\text{H}@r$ and $\text{N}@r$ are Top-$r$ Hit Ratio and Normalized Discounted Cumulative Gain ~\citep{geng2022recommendation,chen2022knowledge}, which are two widely-used metrics for evaluating the performance of LLM-empowered RecSys. 
$\widehat{\text{H}@r}$ and $\widehat{\text{N}@r}$ are the Top-$r$ Hit Ratio and Normalized Discounted Cumulative Gain when the victim model is under attack. 
The larger the decrease in H@$r$ and N@$r$, the better the algorithm's attack performance.
In this paper, $r$ is set to 5 and 10, respectively. 
For \textbf{TALLRec} model, the recommendation results only contain "Yes" and "No," which can be considered as a binary classification task. We adopt Area Under the Receiver Operating Characteristic (AUC) as the metric to measure the recommendation performance, which is consistent with the work of ~\citet{bao2023tallrec}. 
$\text{ASR-A}=1-{\widehat{\text{AUC}}}/{\text{AUC}}$ is introduced to evaluate the attack performance, where ${\widehat{\text{AUC}}}$ is the AUC when the TALLRec is under attacks.

\subsection{Attack Effectiveness}
\begin{figure}[t]
\centering
\subfigure[AUC]{
	\label{fig:auc-tallrec}
	\includegraphics[width=1.5in]{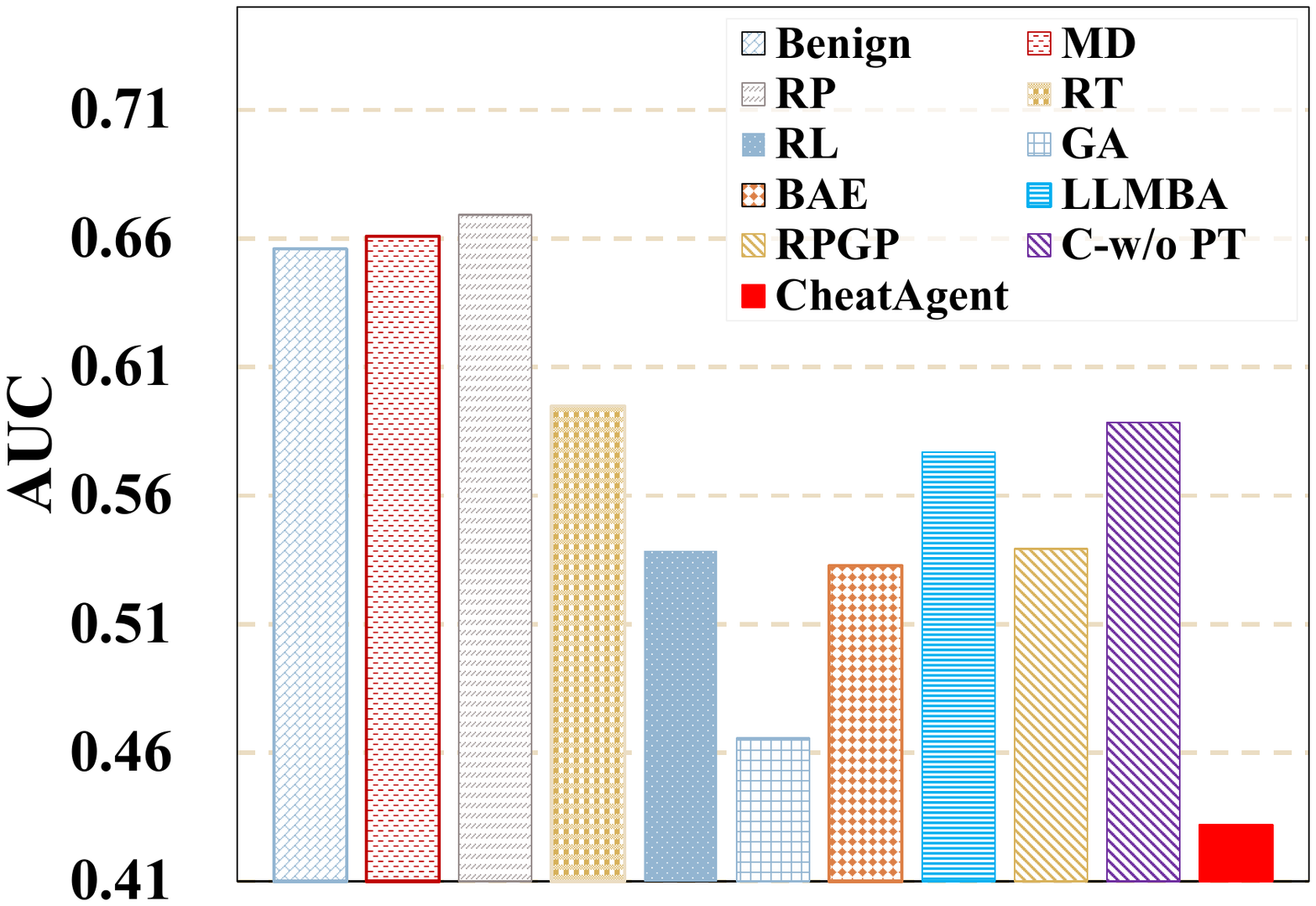}
}
\subfigure[ASR-A]{
	\label{fig:asr-a-tallrec}
	\includegraphics[width=1.5in]{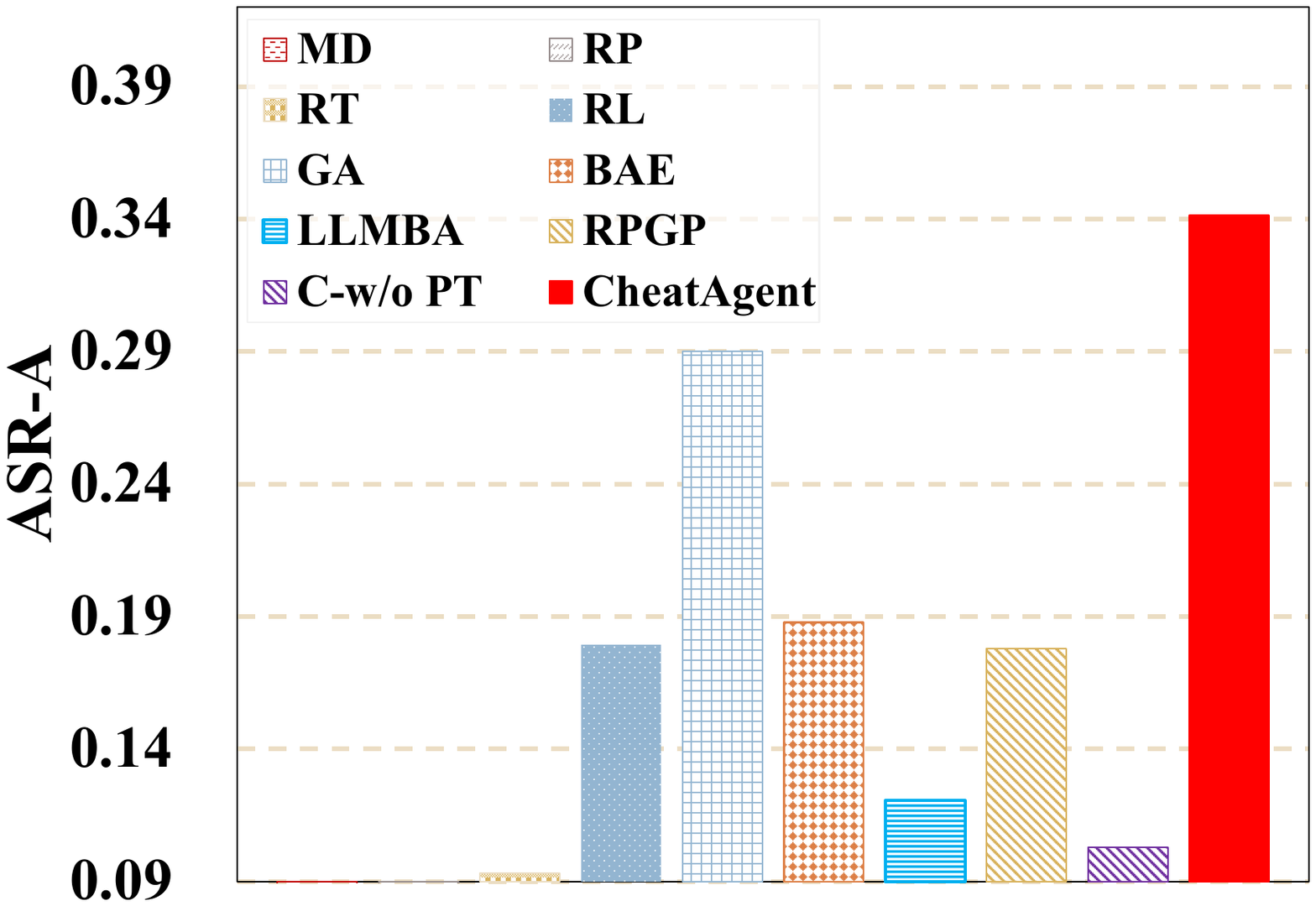}
}
\caption{Attack performance of different methods (Victim model: TALLRec). }
\label{fig:TALLRec}
\end{figure}

We first evaluate the attack effectiveness of the proposed method in this subsection. 
The attack performance of different approaches based on \textbf{P5} are summarised in Table~\ref{tab:p5-sequential} and Table~\ref{tab:p5-random} (\textbf{Appendix}~\ref{appendix:experiments}). 
For \textbf{TALLRec}, the AUC and ASR-A are illustrated in Figure~\ref{fig:TALLRec}. 
Based on comprehensive experiments, we have some following insights:
\begin{itemize}
    \item As shown in Table~\ref{tab:p5-sequential}, the recommendation performance decreases by randomly inserting some token or item perturbations (e.g., RT and RP), indicating that the existing LLM-empowered recommender systems are highly vulnerable. 
    This observation will inspire researchers to pay more attention to the robustness and trustworthiness of utilizing LLMs for other downstream tasks. 
    \item We have discovered that the manually designed adversarial examples, i.e., MD, cannot deceive the target victim model effectively by comparing it with other baselines.  
    Therefore, we require more potent attack strategies instead of relying solely on the manual construction of adversarial examples to explore the vulnerability of LLM-empowered RecSys.
    \item As shown in Table~\ref{tab:p5-sequential} and Table~\ref{tab:p5-random} (\textbf{Appendix}~\ref{appendix:experiments}), the proposed method outperforms other baselines and undermines the recommendation performance dramatically, indicating the effectiveness of the proposed method. 
    Despite the numerous distinctions between P5 and TALLRec, the proposed method effectively deceives both, showcasing its resilience against the architecture of the victim RecSys.
    \item By comparing RPGP with RP and RT, we can observe that inserting random perturbations adjacent to the important tokens leads to a rise in attack performance. 
    This demonstrates the effectiveness of the proposed insertion positioning.
    \item Based on the results of C-w/o PT, we observe that perturbations generated by the LLM-based agent can effectively attack the RecSys even without prompt tuning, demonstrating the potential of the LLM-based agent in performing attacks. 
    Besides, this phenomenon also leads us to speculate that despite the fine-tuning of existing LLM-empowered RecSys on downstream recommendation tasks, they still retain some vulnerabilities of LLMs. 
    \item By comparing the experimental results of C-w/o PT with CheatAgent, we have observed a significant improvement in the attack performance of the agent through policy tuning, demonstrating the effectiveness of the proposed prompt tuning-based attack policy optimization strategy.
\end{itemize}

\begin{figure}[t]
\centering
\subfigure[Cosine similarity]{
	\label{fig:cos}
	\includegraphics[width=1.5in]{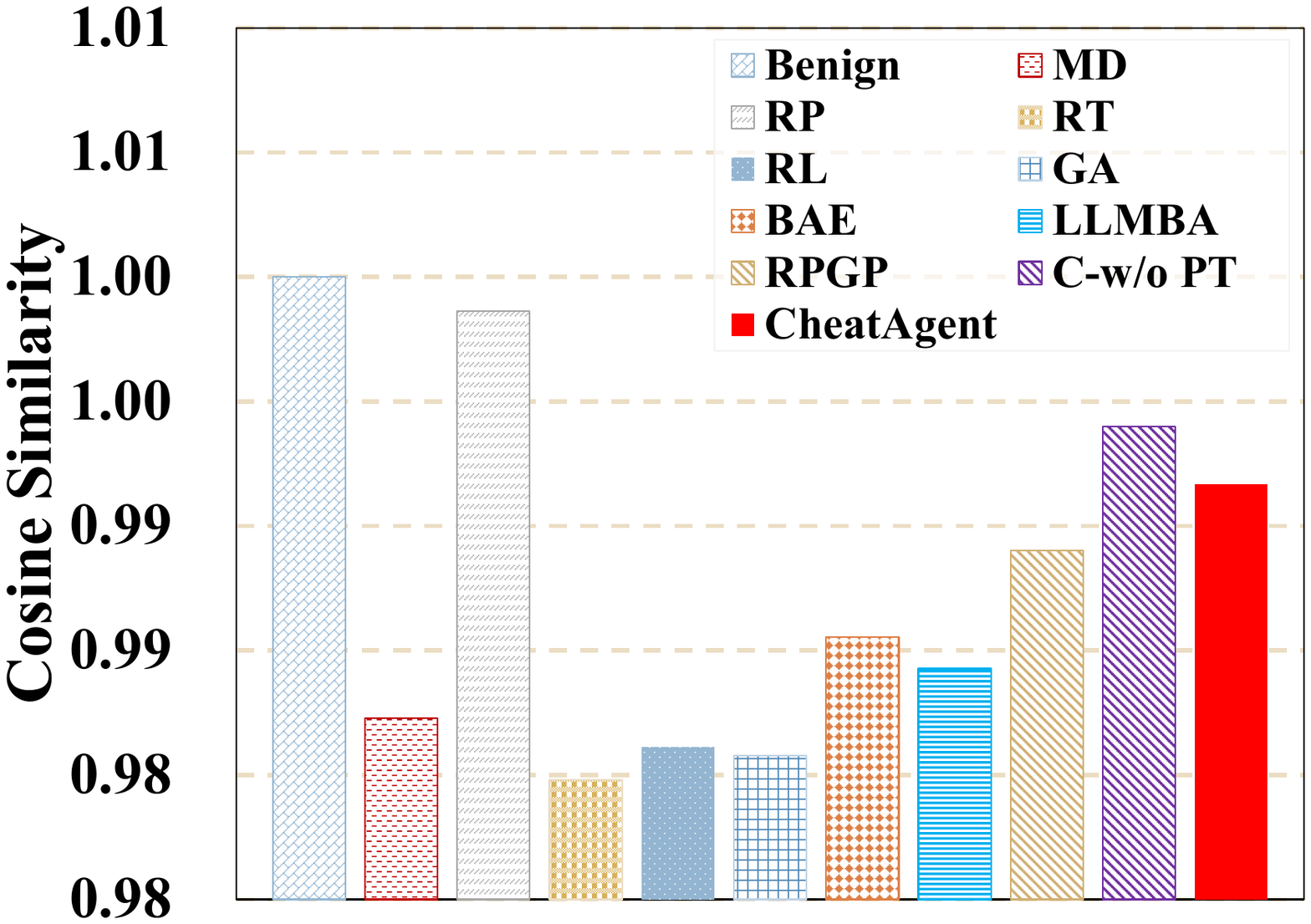}
}
\subfigure[1-Norm]{
	\label{fig:norm}
	\includegraphics[width=1.5in]{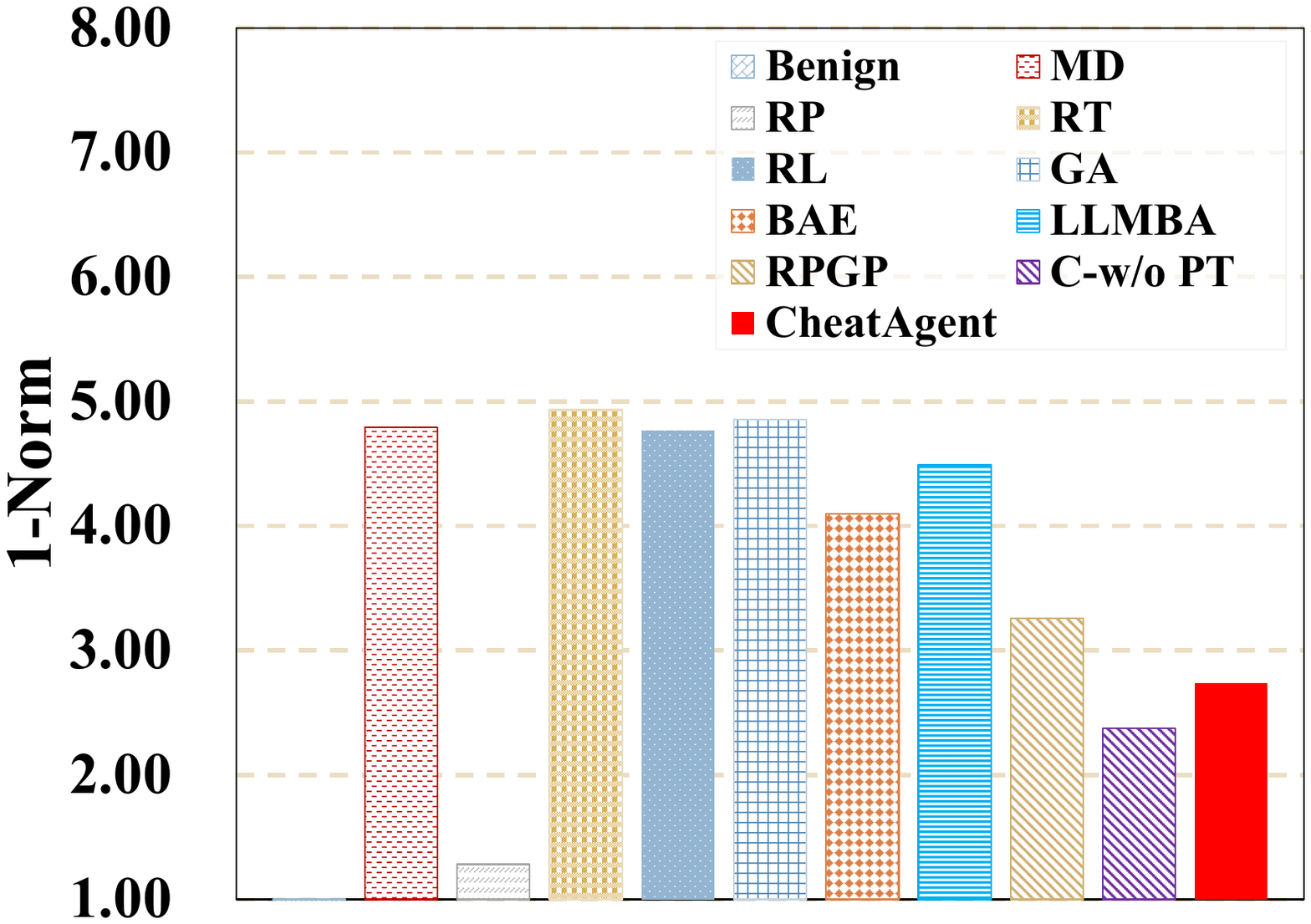}
}
\caption{The semantic similarity between the benign and adversarial prompts.}
\label{fig:semantic_similarity}
\end{figure}

\begin{table}[t]
  \centering
  \caption{Attack Performance of different methods. (Victim Model: P5; Indexing: Sequential)} 
  \scalebox{0.58}{
    \begin{tabular}{|c|c|cccccccc|}
    \toprule
     & Methods & H@5 $\downarrow$ & H@10 $\downarrow$ & N@5 $\downarrow$ & N@10 $\downarrow$ & {\small ASR-H@5} $\uparrow$ & {\small ASR-H@10} $\uparrow$ & {\small ASR-N@5} $\uparrow$ & {\small ASR-N@10} $\uparrow$ \\
    \midrule
    \multirow{11}[3]{*}{\rotatebox{90}{\textbf{ML1M}}} & Benign & 0.2116  & 0.3055  & 0.1436  & 0.1737  & /     & /     & /     & / \\
          & MD    & 0.1982  & 0.2818  & 0.1330  & 0.1602  & 0.0634  & 0.0775  & 0.0735  & 0.0776  \\
          & RP    & 0.2051  & 0.2940  & 0.1386  & 0.1671  & 0.0305  & 0.0374  & 0.0347  & 0.0380  \\
          & RT    & 0.1949  & 0.2800  & 0.1317  & 0.1591  & 0.0790  & 0.0835  & 0.0826  & 0.0839  \\
          & RL    & 0.1917  & 0.2788  & 0.1296  & 0.1576  & 0.0939  & 0.0873  & 0.0974  & 0.0926  \\
          & GA    & \underline{0.0829}  & 0.1419  & 0.0532  & 0.0721  & \underline{0.6080}  & 0.5355  & 0.6298  & 0.5849  \\
          & BAE   & 0.1606  & 0.2440  & 0.1047  & 0.1315  & 0.2410  & 0.2011  & 0.2712  & 0.2432  \\
          & LLMBA & 0.1889  & 0.2825  & 0.1284  & 0.1585  & 0.1072  & 0.0753  & 0.1061  & 0.0876  \\
\cline{2-10}          & RPGP  & 0.1733  & 0.2588  & 0.1164  & 0.1439  & 0.1808  & 0.1528  & 0.1893  & 0.1715  \\
          & C-w/o PT & 0.0844  & \underline{0.1392}  & \underline{0.0531}  & \underline{0.0706}  & 0.6009  & \underline{0.5442}  & \underline{0.6303}  & \underline{0.5935}  \\
          & CheatAgent & \textbf{0.0614} & \textbf{0.1132} & \textbf{0.0389} & \textbf{0.0555} & \textbf{0.7097} & \textbf{0.6293} & \textbf{0.7290} & \textbf{0.6805} \\
      \midrule
    \multirow{11}[3]{*}{\rotatebox{90}{\textbf{LastFM}}} & Benign & 0.0404  & 0.0606  & 0.0265  & 0.0331  & /     & /     & /     & / \\
          & MD    & 0.0339  & 0.0477  & 0.0230  & 0.0274  & 0.1591  & 0.2121  & 0.1333  & 0.1713  \\
          & RP    & 0.0394  & 0.0550  & 0.0241  & 0.0291  & 0.0227  & 0.0909  & 0.0921  & 0.1195  \\
          & RT    & 0.0413  & 0.0550  & 0.0271  & 0.0315  & -0.0227  & 0.0909  & -0.0216  & 0.0463  \\
          & RL    & 0.0294  & 0.0468  & 0.0200  & 0.0256  & 0.2727  & 0.2273  & 0.2460  & 0.2272  \\
          & GA    & 0.0248  & 0.0431  & 0.0156  & 0.0216  & 0.3864  & 0.2879  & 0.4111  & 0.3477  \\
          & BAE   & 0.0165  & 0.0339  & 0.0093  & 0.0149  & 0.5909  & 0.4394  & 0.6480  & 0.5497  \\
          & LLMBA & 0.0404  & 0.0541  & 0.0291  & 0.0336  & 0.0000  & 0.1061  & -0.0969  & -0.0150  \\
\cline{2-10}          & RPGP  & 0.0294  & 0.0514  & 0.0184  & 0.0253  & 0.2727  & 0.1515  & 0.3076  & 0.2349  \\
          & C-w/o PT & \underline{0.0138}  & \underline{0.0275}  & \underline{0.0091}  & \underline{0.0135}  & \underline{0.6591}  & \underline{0.5455}  & \underline{0.6580}  & \underline{0.5924}  \\
          & CheatAgent & \textbf{0.0119} & \textbf{0.0257} & \textbf{0.0072} & \textbf{0.0118} & \textbf{0.7045} & \textbf{0.5758} & \textbf{0.7269} & \textbf{0.6445} \\
    \midrule
    \multirow{11}[4]{*}{\rotatebox{90}{\textbf{Taobao}}} & Benign & 0.1420  & 0.1704  & 0.1100  & 0.1191  & /     & /     & /     & / \\
          & MD    & 0.1365  & 0.1624  & 0.1085  & 0.1170  & 0.0392  & 0.0471  & 0.0130  & 0.0180  \\
          & RP    & 0.1250  & 0.1512  & 0.0977  & 0.1061  & 0.1200  & 0.1125  & 0.1117  & 0.1091  \\
          & RT    & 0.1396  & 0.1658  & 0.1090  & 0.1174  & 0.0173  & 0.0269  & 0.0092  & 0.0145  \\
          & RL    & 0.1376  & 0.1650  & 0.1075  & 0.1163  & 0.0311  & 0.0317  & 0.0222  & 0.0234  \\
          & GA    & 0.1294  & 0.1579  & 0.0993  & 0.1086  & 0.0888  & 0.0731  & 0.0966  & 0.0886  \\
          & BAE   & 0.1278  & 0.1519  & 0.0989  & 0.1066  & 0.1003  & 0.1087  & 0.1009  & 0.1050  \\
          & LLMBA & 0.1353  & 0.1624  & 0.1050  & 0.1138  & 0.0473  & 0.0471  & 0.0452  & 0.0448  \\
\cline{2-10}          & RPGP  & 0.1258  & 0.1512  & 0.0971  & 0.1053  & 0.1142  & 0.1125  & 0.1167  & 0.1159  \\
          & C-w/o PT & \underline{0.1017}  & \underline{0.1258}  & \underline{0.0737}  & \underline{0.0815}  & \underline{0.2837}  & \underline{0.2615}  & \underline{0.3298}  & \underline{0.3161}  \\
          & CheatAgent & \textbf{0.0985} & \textbf{0.1229} & \textbf{0.0717} & \textbf{0.0796} & \textbf{0.3068} & \textbf{0.2788} & \textbf{0.3480} & \textbf{0.3319} \\
    \bottomrule
    \multicolumn{10}{l}{\textbf{Bold} fonts and \underline{underlines} indicate the best and second-best attack performance, respectively.}\\
    \end{tabular}%
    }
  \label{tab:p5-sequential}%
\end{table}%

\subsection{Semantic Similarity}
In this subsection, we test whether inserting adversarial perturbations will change the semantic information of the benign prompt. We use the \emph{bge-large-en} model~\citep{xiao2023c} to map the adversarial and benign prompt to a 512-dimension vector. Cosine similarity and 1-Norm difference are calculated to measure the semantic similarity. 

First, as shown in Figure~\ref{fig:semantic_similarity}, all methods exhibit a high cosine similarity and a low 1-norm difference, primarily due to the imposed constraint on the intensity of perturbations. 
Second, there is a minimal semantic discrepancy between RP and the benign prompt, indicating that inserting perturbations to the users' profiles $V^{u_i}$ is more stealthy than perturbing input prompts $P$. 
Third, apart from RP, our proposed method achieves the highest cosine similarity and the smallest 1-norm difference, demonstrating the effectiveness of our approach in attacking RecSys while maintaining stealthiness. This characteristic makes our method more difficult to detect, thereby posing a greater threat.

\subsection{Ablation Study}
In this subsection, some ablation studies are constructed to investigate the effectiveness of each proposed component. 
Three variants are introduced here for comparison: 1) CheatAgent-RP uses the LLM agent-empowered perturbation
generation to produce perturbations and insert them into the random positions. 2) CheatAgent-I fine-tunes the prefix prompt with random initialization. 3) CheatAgent-T directly employs the initial prefix prompt to produce the adversarial perturbations without further policy tuning. 
The results are shown in Table~\ref{tab:ablation_exp}. 
Through the comparison of CheatAgent with CheatAgent-RP, we demonstrate that the insertion of perturbations into random positions within the input leads to a significant decrease in attack performance. Therefore, it is imperative to identify the token with the maximum impact in order to enhance the attack success rate. 
By comparing the results of CheatAgent with those of CheatAgent-I and CheatAgent-T, we demonstrate that both the initial policy generation and the self-reflection policy optimization processes are necessary for the LLM-based agent to increase the attack performance. 

\begin{table*}[htbp]
  \centering
  \caption{Comparison between CheatAgent and its variants on three datasets. Bold fonts denotes the best performance.}
  \scalebox{0.7}{
    \begin{tabular}{cccccccccc}
    \toprule
    Datasets & Methods & H@5 $\downarrow$ & H@10 $\downarrow$ & N@5 $\downarrow$ & N@10 $\downarrow$ & ASR-H@5 $\uparrow$ & ASR-H@10 $\uparrow$ & ASR-N@5 $\uparrow$ & ASR-N@10 $\uparrow$ \\
    \midrule
    \multirow{4}[2]{*}{LastFM} 
    & CheatAgent & \textbf{0.0119} & \textbf{0.0257} & \textbf{0.0072} & \textbf{0.0118} & \textbf{0.7045} & \textbf{0.5758} & \textbf{0.7269} & \textbf{0.6445} \\
          & CheatAgent-RP & 0.0193  & 0.0358  & 0.0111  & 0.0166  & 0.5227  & 0.4091  & 0.5816  & 0.4995  \\
          & CheatAgent-I & 0.0147  & 0.0284  & 0.0096  & 0.0140  & 0.6364  & 0.5303  & 0.6377  & 0.5769  \\
          & CheatAgent-T & 0.0128  & 0.0259  & 0.0074  & 0.0120  & 0.6818  & 0.5730  & 0.7199  & 0.6371  \\
    \midrule
    \multirow{4}[2]{*}{ML1M} & CheatAgent & \textbf{0.0614} & \textbf{0.1132} & \textbf{0.0389} & \textbf{0.0555} & \textbf{0.7097} & \textbf{0.6293} & \textbf{0.7290} & \textbf{0.6805} \\
          & CheatAgent-RP & 0.1336  & 0.2036  & 0.0881  & 0.1107  & 0.3685  & 0.3333  & 0.3866  & 0.3630  \\
          & CheatAgent-I & 0.0810  & 0.1354  & 0.0512  & 0.0686  & 0.6174  & 0.5566  & 0.6437  & 0.6050  \\
          & CheatAgent-T & 0.0727  & 0.1205  & 0.0456  & 0.0608  & 0.6565  & 0.6054  & 0.6825  & 0.6497  \\
    \midrule
    \multirow{4}[2]{*}{Taobao} & CheatAgent & \textbf{0.0985} & \textbf{0.1229} & \textbf{0.0717} & \textbf{0.0796} & \textbf{0.3068} & \textbf{0.2788} & \textbf{0.3480} & \textbf{0.3319} \\
          & CheatAgent-RP & 0.1258  & 0.1497  & 0.0960  & 0.1037  & 0.1142  & 0.1212  & 0.1271  & 0.1293  \\
          & CheatAgent-I & 0.1024  & 0.1263  & 0.0744  & 0.0821  & 0.2791  & 0.2587  & 0.3233  & 0.3107  \\
          & CheatAgent-T & 0.0985  & 0.1243  & 0.0718  & 0.0802  & 0.3068  & 0.2702  & 0.3468  & 0.3272  \\
    \bottomrule
    \end{tabular}%
    }
  \label{tab:ablation_exp}%
\end{table*}%

\subsection{Parameter Analysis}
\begin{figure}[t]
\centering
\subfigure[$\text{H}@r$ and $\text{N}@r$ w.r.t. $k$]{
	\includegraphics[width=1.45in]{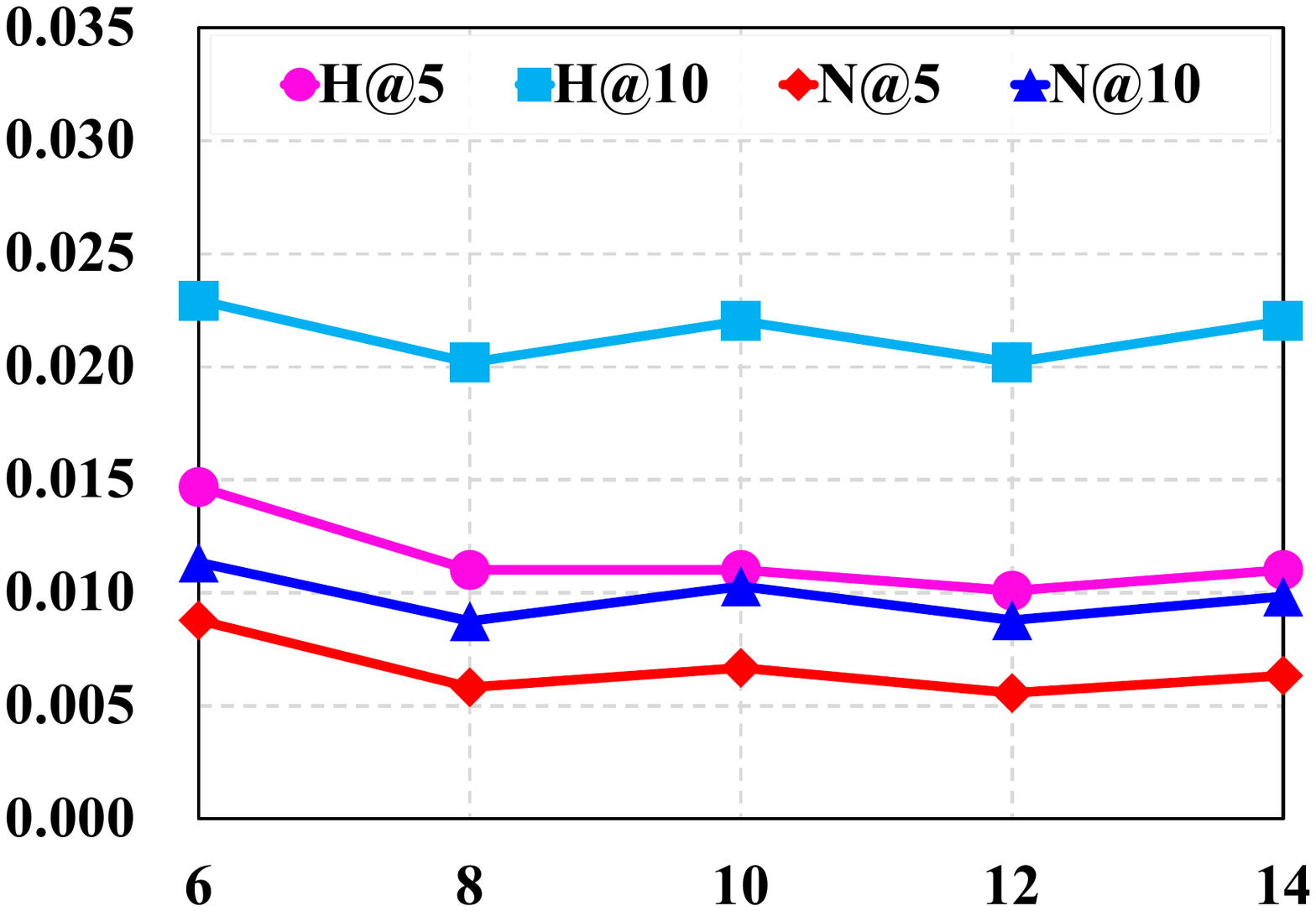}
}
\subfigure[$\text{ASR-A}@r$ and $\text{ASR-N}@r$ w.r.t. $k$]{
	\includegraphics[width=1.45in]{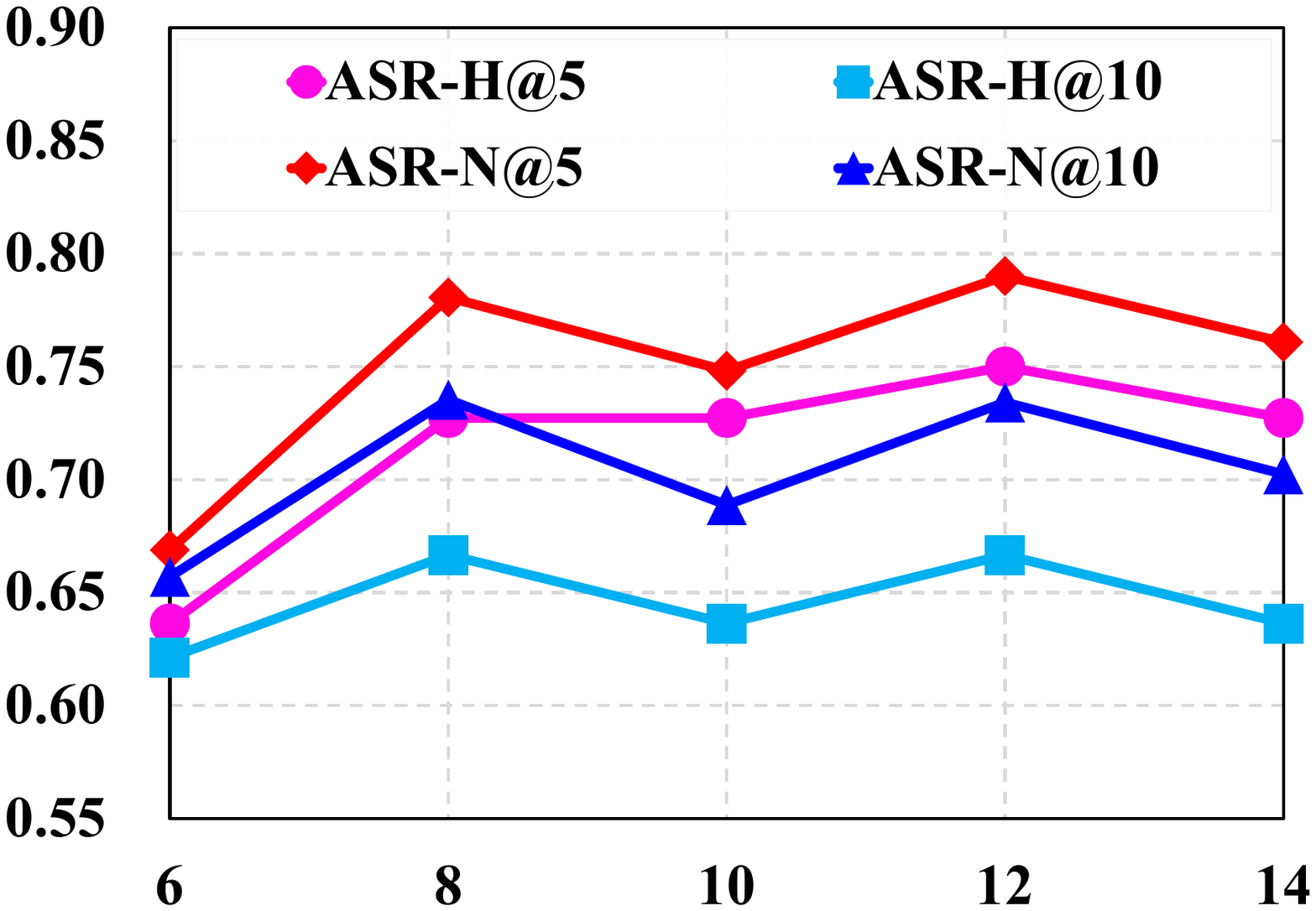}
}
\subfigure[$\text{H}@r$ and $\text{N}@r$ w.r.t. $n$]{
	\includegraphics[width=1.45in]{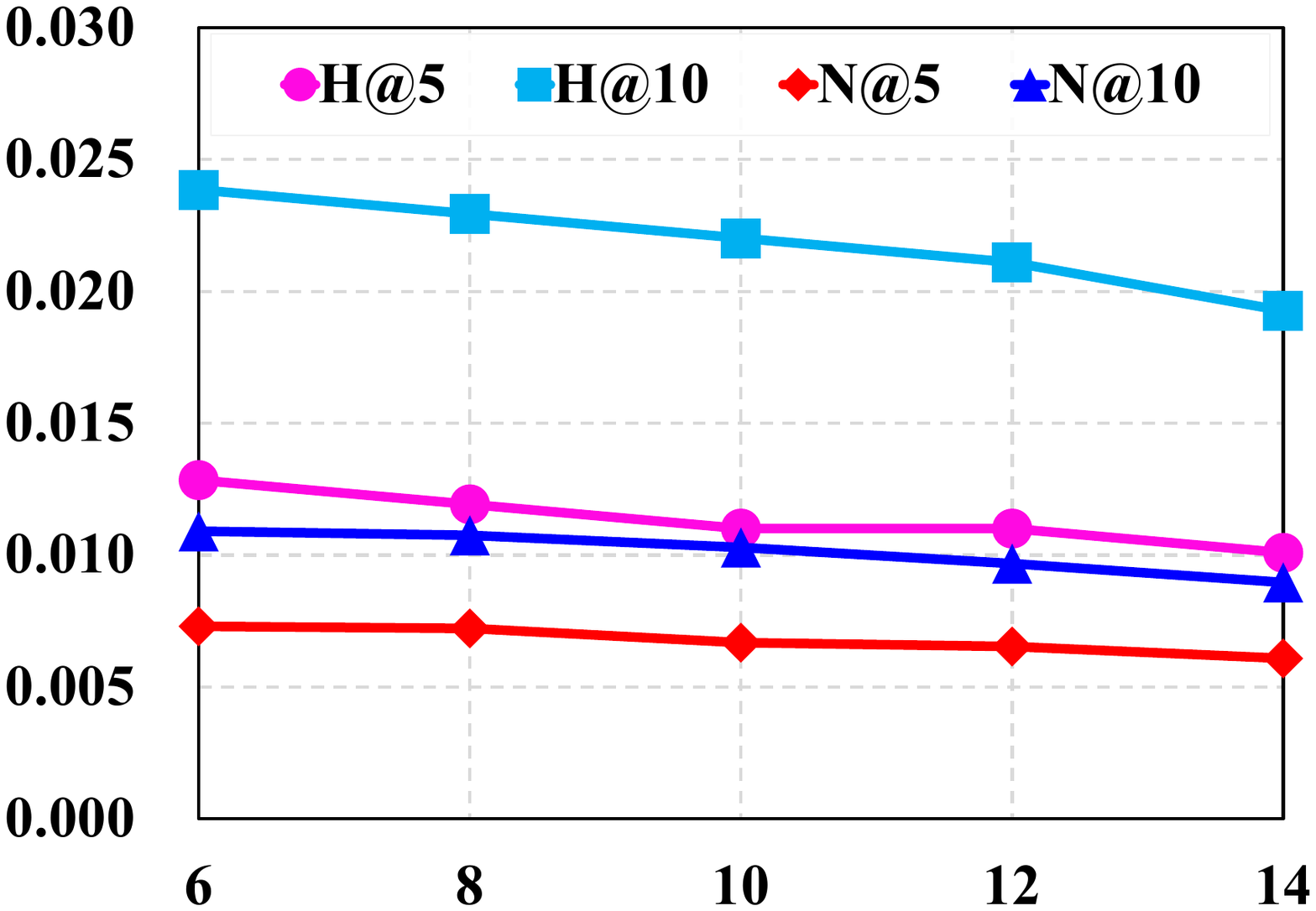}
}
\subfigure[$\text{ASR-A}@r$ and $\text{ASR-N}@r$ w.r.t. $n$]{
	\includegraphics[width=1.45in]{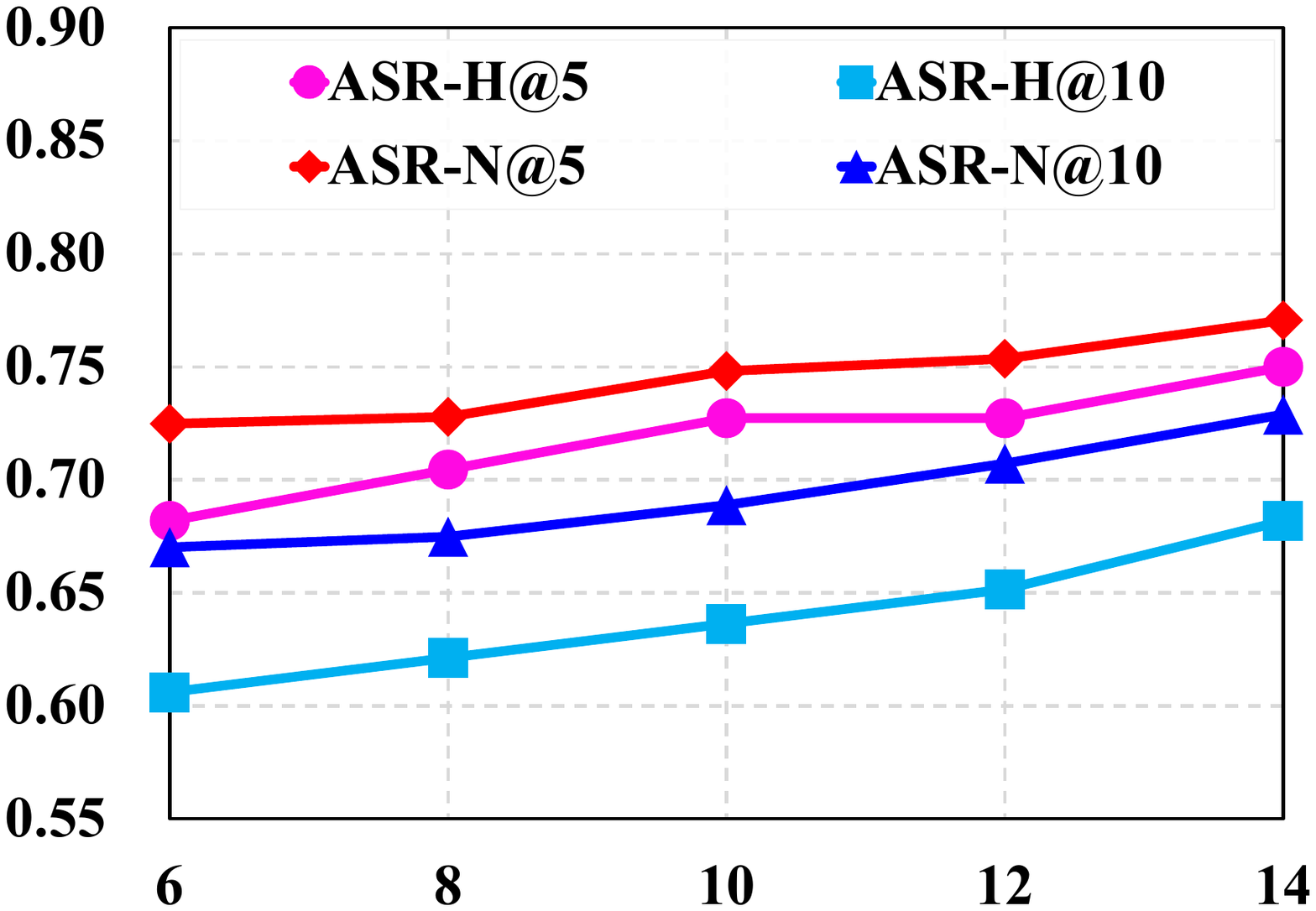}
}
\caption{Effect of the hyper-parameters $k$ and $n$. }
\label{fig:parameter_analysis}
\end{figure}

In this subsection, we study the impact of model hyper-parameters. There are mainly two hyper-parameters, i.e., $n$ and $k$, associated with the attack performance. 
$k$ is the number of the randomly initialized prefix prompt during the initial policy generation process. Given an attack instruction, $n$ is the number of the generated perturbations of the LLM-based agent. 
We fix one of them and gradually vary the other, observing its impact on the attack performance. 
The results are illustrated in Figure~\ref{fig:parameter_analysis}. 
With the change of $k$, the $\text{H}@r$, $\text{N}@r$, $\text{ASR-A}@r$ and $\text{ASR-N}@r$ fluctuate within a small range, which demonstrates the robustness of the proposed method to the hyper-parameters $k$. 
As for $n$, the attack performance gradually strengthens as $n$ increases. However, large $n$ will consume abundant time. Consequently, we set $n=10$ as the default in this paper to achieve a balance of the attack performance and efficiency.

\section{Related Work}\label{sec:related_work}
In this section, we briefly overview some related studies focusing on adversarial attacks for recommender systems. 
Due to the space limitation, some studies about the LLM-empowered RecSys and vulnerabilities of LLM are reviewed in \textbf{Appendix}~\ref{appendix:related_work}.

Generally, adversarial attacks for recommender systems are broadly divided into two categories~\cite{fan2022comprehensive}: 1) \textbf{Evasion Attack} happens during the inference phase. Given a fixed, well-trained RecSys, attackers aim to modify the user's profiles to manipulate the recommendation outcome. 2) \textbf{Poisoning Attack} occurs during the data collection before model training. The attackers inject the poisoned fake users into the training set to misguide the model training and undermine its overall performance.

Early methods including heuristic attacks~\cite{williams2006profile,burke2005limited} and gradient-based attacks~\cite{li2016data,christakopoulou2019adversarial} have demonstrated a high rate of success in attacking white-box recommendation models. However, these methods cannot be directly applied to attack black-box recommender systems (RecSys) due to the limited knowledge about the victim model.
Recently, reinforcement learning has emerged as a viable approach for attacking the black-box victim model.
PoisonRec is the first black-box attack framework, which leverages the reinforcement learning architecture to automatically learn effective attack strategies~\cite{song2020poisonrec}.
Chen et al.~\cite{chen2022knowledge} propose a knowledge-enhanced black-box attack by exploiting items' attribute features (i.e., Knowledge Graph) to enhance the item sampling process.
Instead of generating fake users' profiles from scratch, Fan et al.~\cite{fan2021attacking,fan2023adversarial} have developed a copy-based mechanism to obtain real user profiles for poisoning the target black-box RecSys. 
MultiAttack~\cite{fan2023untargeted} also considers utilizing social relationships to degrade the performance of RecSys.

\section{Conclusion}
In this paper, we propose a novel attack framework \textbf{CheatAgent} by introducing an autonomous LLM agent to attack LLM-empowered recommender systems under the black-box scenario.
Specifically, our method first identifies the insertion position for maximum impact with minimal input modification.
Subsequently, CheatAgent crafts subtle perturbations to insert into the prompt by leveraging the LLM as the attack agent.
To improve the quality of adversarial perturbations, we further develop prompt tuning techniques to improve attacking strategies via feedback from the victim RecSys iteratively.
Comprehensive experiments on three real-world datasets show the effectiveness of our proposed methods and highlight the vulnerability of LLM-empowered recommender systems against adversarial attacks.

\begin{acks}
The research described in this paper has been partly supported by the National Natural Science Foundation of China (project no. 62102335), General Research Funds from the Hong Kong Research Grants Council (project no. PolyU 15200021, 15207322, and 15200023), internal research funds from The Hong Kong Polytechnic University (project no. P0036200, P0042693, P0048625, P0048752, and P0051361), Research Collaborative Project no. P0041282, and SHTM Interdisciplinary Large Grant (project no. P0043302). 
\end{acks}

\bibliographystyle{ACM-Reference-Format}
\bibliography{Reference}

\clearpage


\appendix

\section{Whole process of CheatAgent}\label{appendix:pseudo}

\begin{algorithm}[htbp]
  \caption{\textbf{CheatAgent}}  
  \label{al:CheatRec}
  \KwIn{\\
Input $X$, LLM agent $\mathcal{A}$, Attacker's Instruction $\mathcal{P} \in \{ \mathcal{P}_P, \mathcal{P}_{V^{u_i}} \}$, iteration $T$.\\
\textbf{Output:} Adversarial perturbations ${\hat \delta}_T$.\\
\textbf{Procedure:}}
Mask each token within $X$ and find the tokens $\mathcal{S}$ with maximal impact for perturbation insertion \;
\For{$s_i$ \text{in} $\mathcal{S}$}
{
Randomly initialize $k$ prefix prompts $[\mathcal{F}_1,...,\mathcal{F}_k]$ \;
Generate perturbation candidates $\mathcal{B}_j, j \in \{1, k\}$ according to Eq~\eqref{eq:initialization_candidate_generation} \;
Select the optimal initialization of the prefix prompt $\mathcal{F}_0$ according to Eq~\eqref{eq:selection} \;

\For{t \text{in} 1:T}
{
Generate a set of perturbations $\mathcal{B}_T$ \;
Divide the perturbation into positive and negative categories according to Eq~\eqref{eq:selector} \;
Compute the loss according to Eq~\eqref{eq:prefix} \;
Update the prefix prompt according to $\mathcal{F}_{T+1} =  \mathcal{F}_{T} - \gamma \cdot \nabla_{\mathcal{F}_{T}} \mathcal{L}_{\mathcal{F}_{T}}$ \;
}
}
Select the optimal perturbation ${\hat \delta}_T$ according to Eq~\eqref{eq:final_perturbation} \;
\textbf{end for}\\
\end{algorithm}

\section{Experimental Details}\label{appendix:experimental_details}
Due to the space limitation, some details of the experiments and discussions are shown in this section. 

\subsection{Datasets Statistics}\label{appendix:datasets}
We utilize three datasets, i.e., \textbf{ML1M}, \textbf{LastFM}, and \textbf{Taobao}, to construct comprehensive experiments. 
The ML1M dataset is a widely-used benchmark dataset in the field of recommender systems, which contains rating data from the MovieLens website, specifically collected from around 6,040 users and their interactions with around 3,000 movies. The dataset provides information such as user ratings, movie attributes, and timestamps, making it suitable for various recommendation tasks and evaluation of recommendation algorithms. 
The LastFM dataset is another popular dataset, which consists of user listening histories from the Last.fm music streaming service. The dataset includes information about user listening sessions, such as artist and track names, timestamps, and user profiles. 
The Taobao dataset is a large-scale e-commerce dataset collected from the Taobao online shopping platform. It contains a rich set of user behaviors, including browsing, searching, clicking, and purchasing activities. The dataset provides valuable insights into user preferences, purchasing patterns, and item characteristics. 

For \textbf{P5} model, all used datasets are processed according to the work of ~\citet{geng2022recommendation,xu2023openp5}. 
For \textbf{TALLRec} model, we process the \textbf{ML1M} dataset according to the work of ~\citet{bao2023tallrec}. It should be noted that TALLRec divides the users' profiles with extensive interactions into multiple segments, resulting in numerous similar users with only one or two different items in their profiles. 
To be more efficient, we randomly select 1,000 users from the generated datasets to test the performance of different methods.

\subsection{Implementation Details}\label{appendix:implementation}
For \textbf{MD}, we manually design two adversarial prompts to reverse the semantic information of the benign input to guide the victim RecSys to produce opposite recommendations. The manually-designed adversarial prompts are shown in Table~\ref{tab:md}. 
As we mentioned in Section~\ref{sec:promt_tuning}, we use distinct prompts to generate perturbations. The used prompts are shown in Table~\ref{tab:generation_prompt}. 
For \textbf{LLMBA}, we design a similar prompt to generate perturbations, which is also shown in Table~\ref{tab:generation_prompt}.

\subsection{Additional Experiments}\label{appendix:experiments}
\noindent \textbf{Attack Effectiveness. }
Due to the space limitation, the results based on the \textbf{P5} model that uses random indexing strategy are shown in Table~\ref{tab:p5-random}. 
We can observe that, except for the LastFM dataset, the proposed method consistently outperforms other baselines and significantly undermines the recommendation performance.
We argue that the effectiveness of the proposed method on the LastFM dataset is hindered due to the poor recommendation performance of the target RecSys. Consequently, the limited valuable information for policy tuning may impede CheatAgent's attack performance on this dataset. 

\noindent \textbf{Insertion positioning strategy. }
As mentioned in Section~\ref{sec:implementation}, we observe that masking a pair of items and inserting perturbations to the middle of the maximum-impact items can achieve better attack performance. 
To indicate the effectiveness of this strategy, we use a variant of the proposed method for comparison.  
The results are illustrated in Table~\ref{tab:insertion_position}. 
CheatAgent-MI masks each word/item within the input $X$ and inserts perturbations adjacent to the maximum-impact words/item.
From the experiment, we observe that the proposed method outperforms the variant on three datasets, demonstrating the effectiveness of this strategy. 

\begin{table*}[htbp]
  \centering
  \caption{Attack Performance of different methods. We use bold fonts and underlines to indicate the best and second-best attack performance, respectively. (Victim Model: P5; Indexing: Random)}
  \scalebox{0.8}{
    \begin{tabular}{cccccccccc}
    \toprule
    Datasets & Methods & H@5 $\downarrow$ & H@10 $\downarrow$ & N@5 $\downarrow$ & N@10 $\downarrow$ & ASR-H@5 $\uparrow$ & ASR-H@10 $\uparrow$ & ASR-N@5 $\uparrow$ & ASR-N@10 $\uparrow$ \\
    \midrule
    \multirow{11}[4]{*}{ML1M} & Benign & 0.1058  & 0.1533  & 0.0693  & 0.0847  & /     & /     & /     & / \\
          & MD    & 0.0945  & 0.1459  & 0.0619  & 0.0785  & 0.1064  & 0.0486  & 0.1065  & 0.0728  \\
          & RP    & 0.0859  & 0.1320  & 0.0579  & 0.0728  & 0.1878  & 0.1393  & 0.1639  & 0.1401  \\
          & RT    & 0.0901  & 0.1328  & 0.0580  & 0.0718  & 0.1487  & 0.1339  & 0.1631  & 0.1522  \\
          & RL    & 0.0975  & 0.1419  & 0.0648  & 0.0792  & 0.0782  & 0.0745  & 0.0646  & 0.0650  \\
          & GA    & 0.0808  & 0.1248  & 0.0531  & 0.0673  & 0.2363  & 0.1857  & 0.2342  & 0.2046  \\
          & BAE   & 0.0942  & 0.1384  & 0.0611  & 0.0753  & 0.1095  & 0.0972  & 0.1181  & 0.1104  \\
          & LLMBA & 0.0785  & 0.1137  & 0.0528  & 0.0643  & 0.2582  & 0.2581  & 0.2375  & 0.2407  \\
\cmidrule{2-10}          & RPGP  & 0.0783  & 0.1219  & 0.0525  & 0.0665  & 0.2598  & 0.2052  & 0.2420  & 0.2142  \\
          & C-w/o PT & \underline{0.0517}  & \underline{0.0836}  & \underline{0.0329}  & \underline{0.0433}  & \underline{0.5117}  & \underline{0.4546}  & \underline{0.5245}  & \underline{0.4889}  \\
          & CheatAgent & \textbf{0.0449 } & \textbf{0.0742 } & \textbf{0.0283 } & \textbf{0.0377 } & \textbf{0.5759 } & \textbf{0.5162 } & \textbf{0.5923 } & \textbf{0.5546 } \\
    \midrule
    \multirow{11}[3]{*}{LastFM} & Benign & 0.0128  & 0.0248  & 0.0072  & 0.0110  & /     & /     & /     & / \\
          & MD    & 0.0147  & 0.0303  & 0.0078  & 0.0128  & -0.1429  & -0.2222  & -0.0944  & -0.1586  \\
          & RP    & 0.0156  & 0.0229  & 0.0107  & 0.0131  & -0.2143  & 0.0741  & -0.4967  & -0.1867  \\
          & RT    & 0.0092  & 0.0220  & 0.0045  & 0.0087  & 0.2857  & 0.1111  & 0.3678  & 0.2135  \\
          & RL    & \underline{0.0064}  & 0.0174  & 0.0032  & 0.0068  & \underline{0.5000}  & 0.2963  & 0.5501  & 0.3860  \\
          & GA    & 0.0073  & 0.0183  & 0.0038  & 0.0073  & 0.4286  & 0.2593  & 0.4756  & 0.3411  \\
          & BAE   & \textbf{0.0046 } & \textbf{0.0119 } & \textbf{0.0026 } & \textbf{0.0050 } & \textbf{0.6429 } & \textbf{0.5185 } & \textbf{0.6421 } & \textbf{0.5463 } \\
          & LLMBA & 0.0165  & 0.0312  & 0.0094  & 0.0142  & -0.2857  & -0.2593  & -0.3129  & -0.2857  \\
\cmidrule{2-10}          & RPGP  & 0.0119  & 0.0284  & 0.0068  & 0.0121  & 0.0714  & -0.1481  & 0.0496  & -0.0967  \\
          & C-w/o PT & 0.0073  & \underline{0.0174}  & \underline{0.0031}  & \underline{0.0062}  & 0.4286  & \underline{0.2963}  & \underline{0.5687}  & \underline{0.4331}  \\
          & CheatAgent & 0.0101  & 0.0183  & 0.0050  & 0.0075  & 0.2143  & 0.2593  & 0.3067  & 0.3174  \\
          \hline
    \multirow{11}[3]{*}{Taobao} & Benign & 0.1643  & 0.1804  & 0.1277  & 0.1330  & /     & /     & /     & / \\
          & MD    & 0.1584  & 0.1764  & 0.1237  & 0.1296  & 0.0359  & 0.0218  & 0.0315  & 0.0258  \\
          & RP    & 0.1345  & 0.1547  & 0.0983  & 0.1049  & 0.1815  & 0.1426  & 0.2306  & 0.2114  \\
          & RT    & 0.1625  & 0.1797  & 0.1254  & 0.1310  & 0.0110  & 0.0036  & 0.0181  & 0.0149  \\
          & RL    & 0.1609  & 0.1766  & 0.1244  & 0.1296  & 0.0209  & 0.0209  & 0.0259  & 0.0258  \\
          & GA    & 0.1560  & 0.1740  & 0.1189  & 0.1248  & 0.0508  & 0.0354  & 0.0688  & 0.0619  \\
          & BAE   & 0.1517  & 0.1692  & 0.1172  & 0.1229  & 0.0768  & 0.0618  & 0.0827  & 0.0762  \\
          & LLMBA & 0.1592  & 0.1766  & 0.1235  & 0.1291  & 0.0309  & 0.0209  & 0.0330  & 0.0292  \\
\cmidrule{2-10}          & RPGP  & 0.1384  & 0.1563  & 0.1005  & 0.1063  & 0.1575  & 0.1335  & 0.2134  & 0.2007  \\
          & C-w/o PT & \underline{0.1150}  & \underline{0.1378}  & \underline{0.0808}  & \underline{0.0883}  & \underline{0.3001}  & \underline{0.2361} & \underline{0.3671} & \underline{0.3361}  \\
          & CheatAgent & \textbf{0.1071 } & \textbf{0.1317 } & \textbf{0.0742 } & \textbf{0.0823 } & \textbf{0.3480 } & \textbf{0.2698 } & \textbf{0.4189 } & \textbf{0.3815 } \\
    \bottomrule
    \end{tabular}%
    }
  \label{tab:p5-random}%
\end{table*}%

\begin{table*}[htbp]
  \centering
  \caption{Attack performance of different masking methods. (Victim Model: P5; Indexing: Sequential)}
  \scalebox{0.8}{
    \begin{tabular}{cccccccccc}
    \toprule
    Datasets & Methods & H@5 $\downarrow$ & H@10 $\downarrow$ & N@5 $\downarrow$ & N@10 $\downarrow$ & ASR-H@5 $\uparrow$ & ASR-H@10 $\uparrow$ & ASR-N@5 $\uparrow$ & ASR-N@10 $\uparrow$ \\
    \midrule
    \multirow{2}[2]{*}{LastFM} 
    & CheatAgent & \textbf{0.0119} & \textbf{0.0257} & \textbf{0.0072} & \textbf{0.0118} & \textbf{0.7045} & \textbf{0.5758} & \textbf{0.7269} & \textbf{0.6445} \\
          & CheatAgent-MI   & 0.0128  & 0.0259  & 0.0074  & 0.0121  & 0.6818  & 0.5730  & 0.7227  & 0.6352  \\
    \midrule
    \multirow{2}[2]{*}{ML1M} & CheatAgent    & \textbf{0.0614} & \textbf{0.1132} & \textbf{0.0389} & \textbf{0.0555} & \textbf{0.7097} & \textbf{0.6293} & \textbf{0.7290} & \textbf{0.6805} \\
          & CheatAgent-MI    & 0.0697  & 0.1189  & 0.0444  & 0.0603  & 0.6706  & 0.6108  & 0.6908  & 0.6531  \\
    \midrule
    \multirow{2}[2]{*}{Taobao} & CheatAgent    & \textbf{0.0985} & \textbf{0.1229} & \textbf{0.0717} & \textbf{0.0796} & \textbf{0.3068} & \textbf{0.2788} & \textbf{0.3480} & \textbf{0.3319} \\
          & CheatAgent-MI    & 0.1045  & 0.1278  & 0.0760  & 0.0835  & 0.2641  & 0.2500  & 0.3092  & 0.2991  \\
    \bottomrule
    \end{tabular}%
    }
  \label{tab:insertion_position}%
\end{table*}%

\begin{table*}[htbp]
  \centering
  \caption{Manually-designed adversarial prompts.}
  \scalebox{0.7}{
    \begin{tabular}{ccc}
    \toprule
    \multicolumn{1}{c}{Victim Model} & Methods & Used Prompt \\
    \midrule
    \multicolumn{1}{c}{\multirow{2}[4]{*}{P5}} & Benign & What is the top recommended item for {dataset} {user\_id} who interacted with {dataset} item {history}? \\
\cmidrule{3-3}          & MD    & What is \textbf{\red{not}} the top recommended item for {dataset} {user\_id} who interacted with {dataset} item {history}? \\
    \midrule
    \multicolumn{1}{c}{\multirow{2}[4]{*}{TALLRec}} & Benign & Given the user's preference and unpreference, identify whether the user will like the target movie by answering "Yes." or "No.". \\
\cmidrule{3-3}          & MD    & Given the user's preference and unpreference, identify whether the user will \textbf{\red{not}} like the target movie by answering "Yes." or "No.". \\
    \bottomrule
    \end{tabular}%
    }
  \label{tab:md}%
\end{table*}%

\begin{table*}[htbp]
  \centering
  \caption{Used prompt for perturbation generation.}
    \scalebox{0.8}{
  \begin{tabularx}{\textwidth}{l|l|X}
    \toprule
    \multicolumn{2}{c|}{Methods}  & \multicolumn{1}{c}{Used Prompt} \\
    \midrule
    \multirow{3}[1]{*}{CheatAgent} & \multirow{2}[1]{*}{$\mathcal{P}_P$}  & The original input is \textbf{\red{\{Original Prompt\}}} and the prediction is \textbf{\red{\{Target\}}}. Please generate a few letters or words that can change your prediction after inserting it as the \textbf{\red{\{position\}}}-th word of the input. \\
    \cmidrule{2-3}
    & \multirow{1}[1]{*}{$\mathcal{P}_{V^{u_i}}$}  & According to the user's historical sequence \textbf{\red{\{User's Profile\}}}, please predict the items that the user is not interested in. \\
    \hline
    \multicolumn{2}{c|}{\multirow{2}[2]{*}{LLMBA}} & The original input is \textbf{\red{\{Original Prompt\}}} and the prediction is \textbf{\red{\{Target\}}}. Please generate a few letters or words that can change your prediction after inserting it after inserting it at the end of the input sentence. \\
    \bottomrule
    \end{tabularx}%
    }
  \label{tab:generation_prompt}%
\end{table*}%

\section{Related Work}~\label{appendix:related_work}
\subsection{LLM-Empowered Recommender Systems}
The recent breakthrough of LLMs has initiated a new era for RecSys. 
Due to its powerful capability of understanding and reasoning, LLM has been widely used to facilitate various recommendation tasks, such as news recommendation~\cite{wu2023personalized}, drug recommendations~\cite{dongre2023deep}, etc.
For example, BERT4Rec adopts Bidirectional Encoder Representations (i.e., BERT) to model users' sequential behavior for recommendations~\cite{sun2019bert4rec}.
Furthermore, TALLRec aligns the LLM (i.e., LLaMA-7B)  with recommendation data for sequential recommendation~\cite{bao2023tallrec}.
Additionally, by studying the user's historical behavior and preferences, P5 can perform various recommendation tasks such as rating prediction and sequential recommendation and explain the recommendations~\cite{geng2022recommendation}.
In conclusion, LLM-Empowered RecSys is a fast-growing field, and it is necessary to study its vulnerabilities.

\subsection{Adversarial Attack for LLM}
Based on the generation method of adversarial prompts, the existing adversarial attacks for large language models can be broadly divided into three categories: 

\noindent 1) \textbf{Artificial-based Methods.} In the early stage of investigating the vulnerability of LLMs, multiple studies manually devised prompts to bypass the defense mechanism and guide LLMs to generate some harmful content~\cite{kang2023exploiting, wei2023jailbroken}, which is time-consuming and ineffectively with the development of the LLMs.

\noindent 2) \textbf{Optimization-based Methods.} These methods exploit diverse optimization strategies, such as genetic algorithm~\cite{lapid2023open}, gradient-based search~\cite{zou2023universal, zhu2023autodan}, reinforcement learning~\cite{xue2023trojllm}, to find the optimal perturbation. 
For example, \citet{zou2023universal} create the desired adversarial postfix by generating a candidate set according to the gradient and replacing the word from a candidate randomly. 
\citet{lapid2023open} propose to exploit the genetic algorithm to iteratively generate the universal adversarial prompt.

\noindent 3) \textbf{LLM-based Methods.} LLM is employed to generate adversarial samples automatically, which is more efficient and diverse~\cite{deng2023jailbreaker, xu2023instructions}.
\citet{deng2023jailbreaker} propose to exploit the time-based characteristics intrinsic to deconstruct the defense mechanism of LLMs. An automatic method for the generation of adversarial prompts is also presented by fine-tuning the LLM. 
\citet{xu2023instructions} leverage the LLM to generate poisoned instructions and insert the backdoor into LLMs via instruction tuning.

\section{Discussions}~\label{appendix:discussion}

\noindent \textbf{Difference between APRec~\cite{wu2023attacking} and CheatAgent.}
 The objective of APRec~\cite{wu2023attacking} is entirely different from this work. The recommendation model employed by APRec is SASRec~\cite{kang2018self}, which is not a large language model and lacks the ability to comprehend textual language in LLM-based recommendations. Therefore, the vulnerability of LLM-empowered recommender systems is still not explored. To fill the gap in this area, our work takes the pioneering investigation into the vulnerability of LLM-empowered RecSys.

 \noindent \textbf{Practical Applications. }
 The main goal of our research is to investigate the vulnerability of existing LLM-empowered RecSys, so as to spread awareness about the trustworthiness of recommender systems. From the \textbf{industry perspective}, our proposed CheatAgent can assist them in evaluating the vulnerabilities of their deployed LLMs-based recommender systems. The enterprise desires that the LLM-empowered RecSys it employs is robust to small perturbations (e.g., random/bait clicks~\cite{fan2022graph}). Assume that non-English-speaking users who utilize LLM-empowered Shopping Assistant (e.g., Amazon AI Shopping Assistant `Rufus') may unintentionally input their prompts with incorrect singular or plural forms, resulting in an additional character `a', considered as the token perturbation. 
 Alternatively, they may encounter enticing product titles and click on them despite not genuinely liking the products, thereby introducing item perturbation to their history interaction. 
 If such perturbations can significantly impact the recommendation outcomes of LLM-empowered RecSys, leading to the recommendation of undesired products to users, it would undermine their user experience. 
 To prevent such occurrences, the company must investigate the vulnerability of the LLM-empowered RecSys before deploying. 
 In this case, the attacker is the owner (e.g., system manager, system designer, and algorithm developer) of the LLM-empowered RecSys and possesses the ability to access user interaction histories and modify prompts, which is entirely plausible.

Note that the assumptions required for the attack paradigm proposed in this paper are slightly strong since attackers are not always the system's owner and may not be able to manipulate and modify the prompt directly. 
As our work is the first to investigate the vulnerability of LLM-Enpowered RecSys, we believe that the insights presented in this paper can enhance people's attention to the security aspects of the system. 
We also hope that our work can inspire future work to develop more advanced approaches and promote the trustworthiness of LLM-empowered recommender systems.

\noindent \textbf{Query Number and Running Time.}
 We summarize the number of queries and time required to generate an adversarial example for deceiving the victim system, shown as follows:

\begin{table}[htbp]
  \centering
  \caption{Query number and running time of various methods. }
    \scalebox{0.9}{\begin{tabular}{c|c|c}
    \toprule
    Methods & Query Number & Running Time (s) \\
    \hline
    GA    & 550   & 1.22  \\
    BAE   & 151   & 2.72  \\
    RL    & 501   & 5.37  \\
    CheatAgent & 490   & 4.50  \\
    \bottomrule
    \end{tabular}%
    }
  \label{tab:addlabel}%
\end{table}%

Here are some insightful observations from this experiment: 1)
We can observe that the proposed CheatAgent can achieve the best attack performance without significantly increasing the number of queries, demonstrating the effectiveness of the proposed method. Besides, during applications, by leveraging the batch processing capabilities of GPUs/TPUs, we can generate multiple adversarial examples, store them in a list, and feed them into the target system together to significantly decrease the query times.
2) Due to the large action space, the reinforcement learning-based agent (RL) requires more time to generate adversarial examples compared to CheatAgent, which demonstrates the efficiency of the proposed LLM-based agent. 
3) Regarding methods such as GA and BAE, which utilize the genetic algorithm and BERT for perturbation generation, they are faster than the proposed method. The reason is that the proposed CheatAgent introduces an LLM to generate perturbations, which increases the time consumption. However, the discrepancy in running time is marginal and acceptable.


\end{document}